\documentclass[twocolumn]{aa}

\usepackage{natbib,twoopt}
\usepackage{epsfig,times,lscape,rotating}
\usepackage{color}
\usepackage{amsmath}
\usepackage{amssymb}
\usepackage{txfonts}
\usepackage[breaklinks=true]{hyperref} %% to avoid \citeads line fills

 \newcommandtwoopt{\citeads}[3][][]{\href{http://adsabs.harvard.edu/abs/#3}%
                                        {\citealp[#1][#2]{#3}}}
 \newcommandtwoopt{\citepads}[3][][]{\href{http://adsabs.harvard.edu/abs/#3}%
                                        {\citep[#1][#2]{#3}}}
 \newcommandtwoopt{\citetads}[3][][]{\href{http://adsabs.harvard.edu/abs/#3}%
                                        {\citet[#1][#2]{#3}}}
 \newcommandtwoopt{\citealtads}[3][][]{\href{http://adsabs.harvard.edu/abs/#3}%
                                        {\citealt[#1][#2]{#3}}}

\begin{document}

\title{Gaps, rings, and non-axisymmetric structures in protoplanetary disks - from simulations to ALMA observations.}

   \author{M. Flock
          \inst{1}
          \and
	  J.P. Ruge\inst{2}
	  \and
          N. Dzyurkevich\inst{3}
	  \and 
	  Th. Henning\inst{4}
          \and
          H. Klahr\inst{4}
          \and
          S. Wolf\inst{2}
          }

   \institute{CEA UMR AIM Irfu, SAP, CEA-CNRS-Univ. Paris Diderot, Centre de Saclay, F-91191 Gif-sur-Yvette, France\\
              \email{Mario.Flock@cea.fr}
         \and
         Universit\"at zu Kiel, Institut für Theoretische Physik und Astrophysik, Leibnitzstr. 15, 24098 Kiel, Germany            
          \and
          Laboratoire de radioastronomie, UMR 8112 du CNRS, \'{E}cole normale sup\'{e}rieure et Observatoire de Paris, 24 rue Lhomond, F-75231 Paris Cedex 05, France
          \and
	  Max Planck Institute for Astronomy, K\"onigstuhl 17, 69117 Heidelberg, Germany\\   
             }

   \date{}

   \abstract
{}
% Aims
{Recent observations by the Atacama Large Millimeter/submillimeter Array (ALMA) of disks around young stars revealed distinct asymmetries in the dust continuum emission. In this work we wish to study axisymmetric and non-axisymmetric structures, that are generated by the magneto-rotational instability in the outer regions of protoplanetary disks. 
We combine the results of state-of-the-art numerical simulations with post-processing radiative transfer (RT) to generate synthetic maps and predictions for ALMA.}
%Methods
{We performed non-ideal global 3D magneto-hydrodynamic (MHD) stratified simulations of the dead-zone outer edge using the FARGO MHD code PLUTO. The stellar and disk parameters were taken from a parameterized disk model applied for fitting high-angular resolution multi-wavelength observations of various circumstellar disks. We considered a stellar mass of $\rm M_*=0.5 M_\sun$ and a total disk mass of about $\rm 0.085 M_*$. The 2D initial temperature and density profiles were calculated consistently from a given surface density profile and Monte Carlo radiative transfer. The 2D Ohmic resistivity profile was calculated using a dust chemistry model. We considered two values for the dust--to--gas mass ratio, $10^{-2}$ and $10^{-4}$, which resulted in two different levels of magnetic coupling. The initial magnetic field was a vertical net flux field. The radiative transfer simulations were performed with the Monte Carlo-based 3D continuum RT code MC3D. The resulting dust reemission provided the basis for the simulation of observations with ALMA.}
%Results
{All models quickly turned into a turbulent state. The fiducial model with a dust--to--gas mass ratio of $10^{-2}$ developed a large gap followed by a jump in surface density located at the dead-zone outer edge. The jump in density and pressure was strong enough to stop the radial drift of particles at this location. In addition, we observed the generation of vortices by the Rossby wave instability at the jump location close to 60 AU. The vortices were steadily generated and destroyed at a cycle of 40 local orbits. The RT results and simulated ALMA observations predict that it is feasible to observe these large-scale structures that appear in magnetized disks without planets. 
%We adapt our models on the system IRS 48 and show the possibility to explain the famous observations of the major asymmetric inside the disk. 
Neither the turbulent fluctuations in the disk nor specific times of the model can be distinguished on the basis of high-angular resolution submillimeter observations alone. The same applies to the distinction between gaps at the dead-zone edges and planetary gaps, to the distinction between turbulent and simple unperturbed disks, and to the asymmetry created by the vortex.}
{}

\keywords{accretion discs, magnetohydrodynamics (MHD)}
\titlerunning{Gaps, rings, and non-axisymmetric structures in protoplanetary disks}

\maketitle

\section{Introduction}
Observations with the Atacama Large Millimeter/submillimeter Array (ALMA) of nearby young stars are able, for the first time, to resolve the detailed disk structures in the dust continuum emission. Recently, a number of observations showed large asymmetry at submillimeter wavelengths, for instance in the systems Oph IRS 48 \citep{van13}, LkH$\alpha$ 330 \citep{ise13}, SAO 206462 and SR 21 \citep{per14}, or in HD 142527 \citep{cas13,fug13}, in which one has recently found a low-mass stellar companion \citep{bet12,clo14}. Another characteristic feature of such disks is that their inner parts are depleted of dust \citep{bro09}, which is possibly caused by particle traps in the outer regions. These particle traps and concentrations are in general modeled by a jump in the surface density. The increase of density and pressure at the rise of the jump can remove the radial pressure gradient and so lead to Keplerian-rotating gas that stops the radial drift of particles, and leading to a ring structure \citep{mer14}. Another way to create dust rings by a two-component secular gravitational instability was presented recently by \citet{tak14}. At the same time, a jump in surface density can trigger the formation of vortices by the Rossby wave instability (RWI), which in turn leads to a concentration of particles \citep{bar95,wol02,joh04,var06,ina06,lyr08,lyr09,lyr09b,reg12,meh12}.
%
%Such asymmetries are in general modeled by particle concentration in vortices of different strength located at the outer disk \citep{lyr13}. 
But while the trapping mechanism of vortices is well understood \citep{bar95,kla97,meh12,bir13,lyr13,zhu14a}, it is still unclear which process triggers their formation. In most cases, it has been assumed that an embedded planet inside the disk causes perturbations of the surface density, leading to the formation of a vortex and following particle concentration \citep{kol03,dev07,meh10,meh12b,lin12,ata13,zhu14a}. Another possibility was presented by \citet{var06}, \citet{lyr09}, \citet{reg12}, and \citet{fau14}: the surface density enhancement and following vortex formation was generated by the increase of accretion stress in the outer disk caused by the magneto-rotational instability (MRI) \citep{bal91,bal98}. At this location, models predict an increase of accretion stress from a low-ionization region, the so-called dead zone, in which the MRI activity is suppressed \citep{bla94,san02a,tur10,flo12}, to a zone with highly ionized gas in which the MRI is active \citep{dzy13,tur14}. In contrast to models of the inner dead-zone edge, in which a sharp jump in accretion stress is predicted \citep{dzy10,lyr12}, recent models of the outer dead-zone edge present a very smooth transition from low to high gas ionization \citep{dzy13,tur14}. Recent unstratified simulations by \citep{lyr14} showed that the RWI could operate there as well, even with a smooth transition in ionization.\\
\begin{figure*}[t]
\hspace{-1cm}
\begin{minipage}{0.4\textwidth}
\psfig{figure=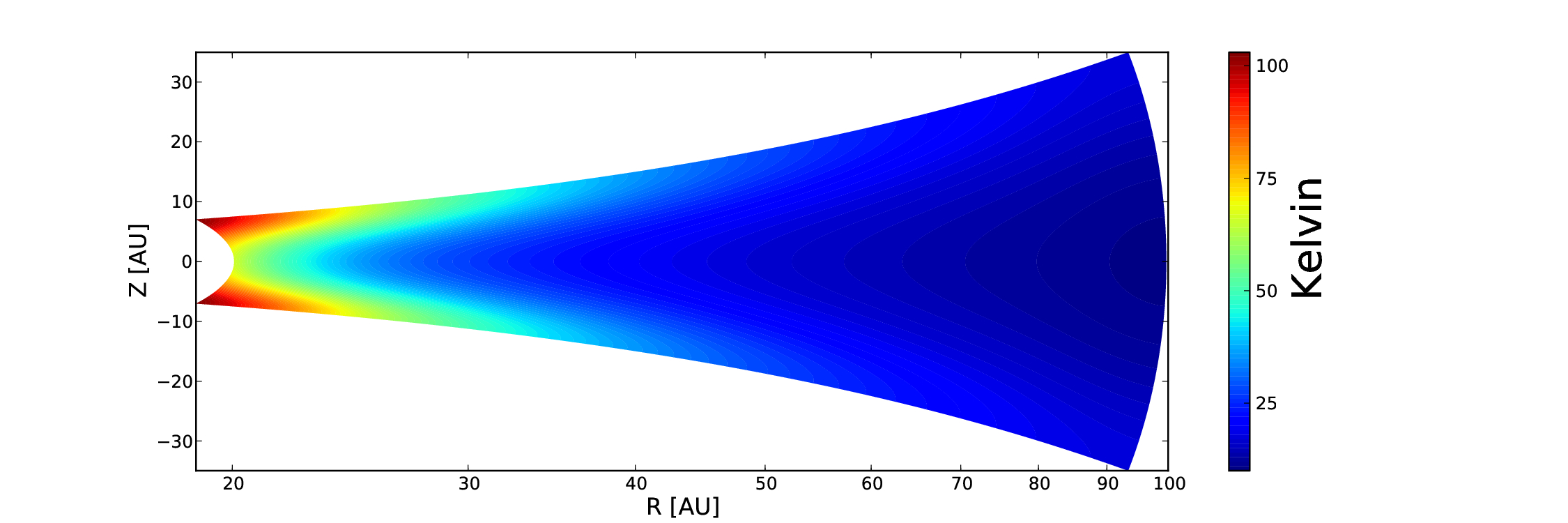,scale=0.30}
\psfig{figure=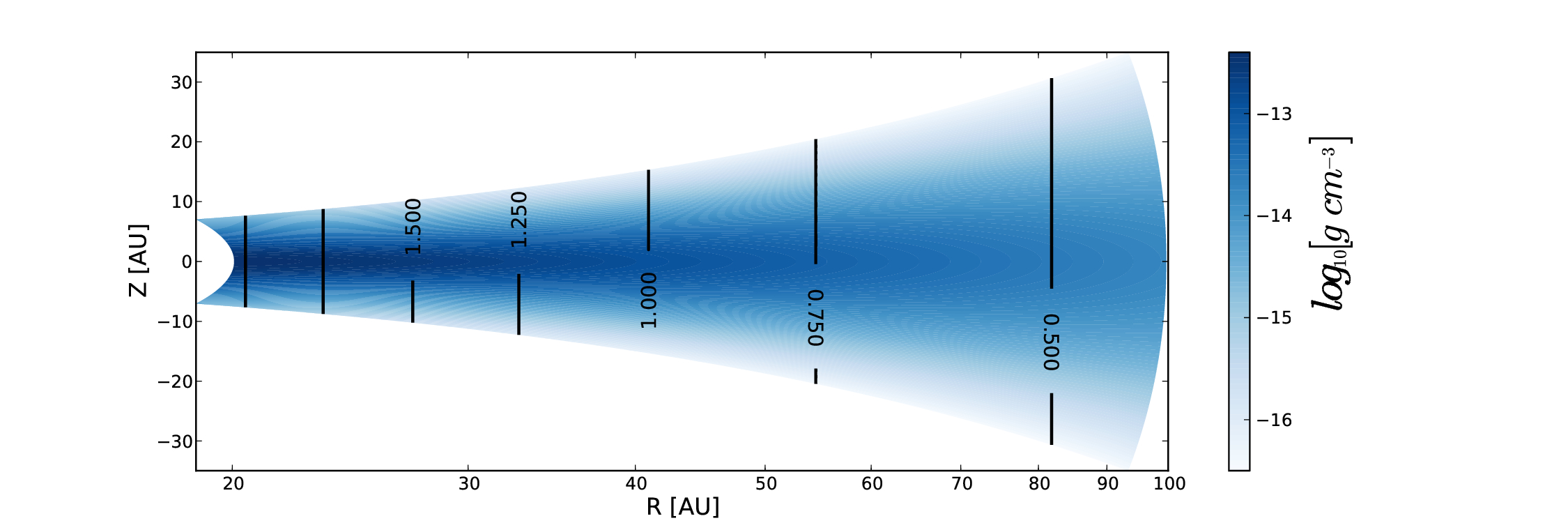,scale=0.30}
\end{minipage}
\hspace{2.5cm}
\begin{minipage}{0.4\textwidth}
\psfig{figure=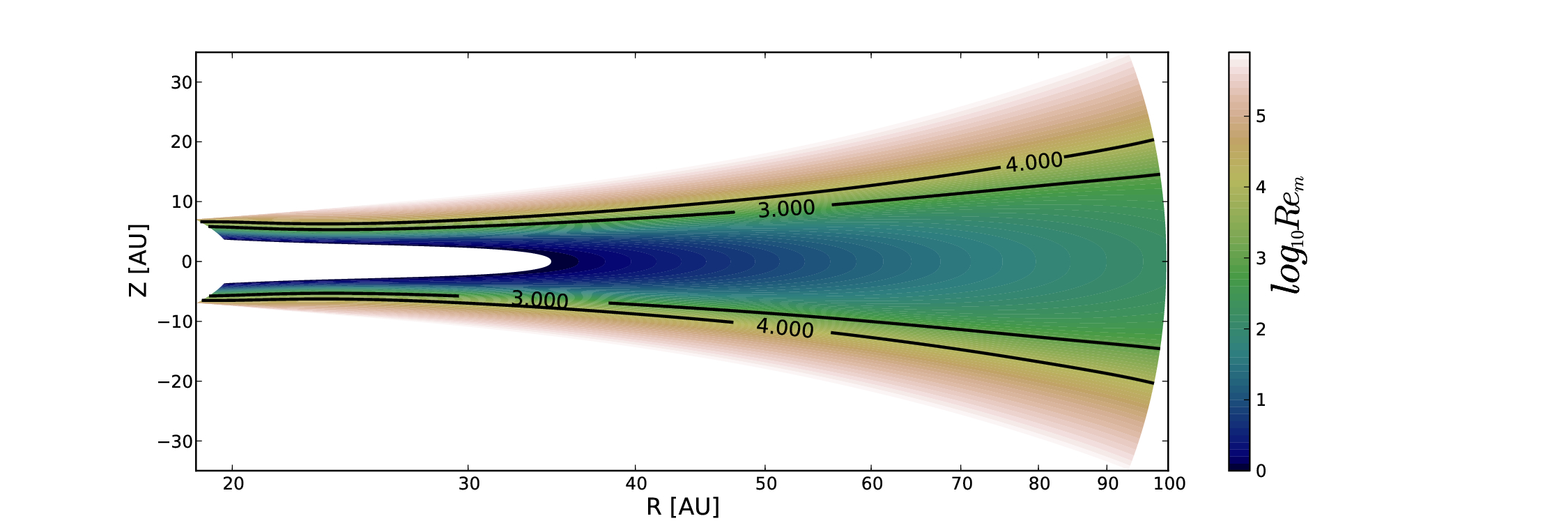,scale=0.30}
\psfig{figure=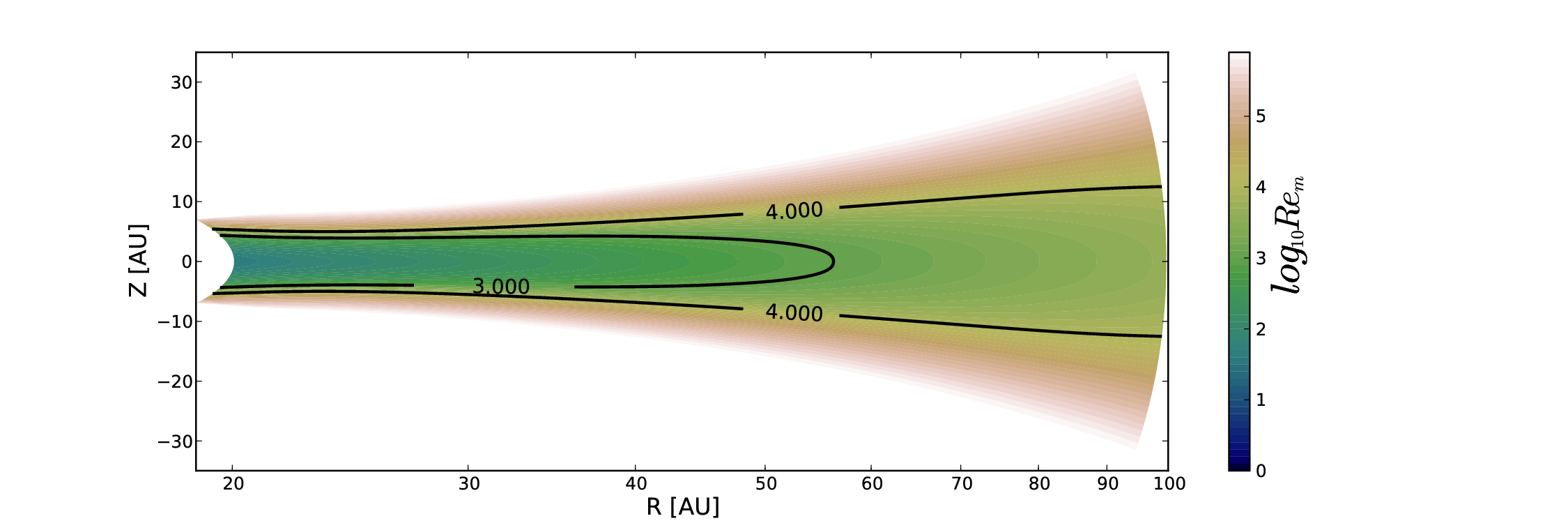,scale=0.30}
\end{minipage}
\caption{Initial temperature distribution ({\it top left}) and logarithmic density distribution ({\it bottom left}). The black contour lines show the strength of the initial vertical magnetic field in steps of $\rm 0.25$ mGauss, starting with 2 mGauss close to 20 AU. Initial resistivity profile with a dust--to--gas mass ratio of $10^{-2}$ ({\it top right}) and $10^{-4}$ ({\it bottom right}). Overplotted are contour lines of constant magnetic Reynolds number $\rm Re_m=c_s H/\eta$ with $\rm Re_m=10^3$ and $10^4$.}
\label{fig:INIT}
\end{figure*}
%Models of particle concentration usually assume a perturbation in the radial pressure profile leading to a local region of Keplerian or even super-Keplerian gas rotation inside the disk. A processes leading to a long lived pressure bump called Zonal Flow was found in local \citep{joh09,dit13,joh14} and global \citep{uri11} simulations in MHD turbulent disks and recently also in local non-ideal MHD simulations \citep{sim14}. Dependent on the strength and lifetime of the zonal flow it is possible to stop or reduce the radial drift of partially decoupled particles. While the formation and trapping mechanism of vortices is well understood \citep{kla97,meh12} it is still unclear which process triggers their formation as they need special conditions to be generated. One possible location for vortex formation and particle trapping is believed to be the inner dead-zone edge of protoplanetary disks where a jump in accretion stress create a surface density enhancement \citep{dzy10,lyr12}.
%In our work we study the structure formation triggered due to the magneto-rotational instability (MRI) \citep{bal91,bal98} with focus on to the outer region (20 to 100 AU) of a T Tauri protoplanetary disks where there is a transition between low and high magnetic gas coupling due to the different levels of ionization \citep{dzy13,tur14}. Depending on the ionization degree, the MRI-generated turbulence will be reduced to a low turbulence regime called the dead zone in which the MRI activity is suppressed \citep{bla94,fle00,san02a,inu05,tur07,tur10,flo12}.
%
In this work we investigate the gas dynamics and structure formation in the outer regions of magnetized protoplanetary disks by combining the latest global simulations with accurate initial conditions obtained by physical constraints from observations, radiative transfer, and chemical modeling. 
The 3D non-ideal global simulations are performed with the FARGO MHD code PLUTO \citep{mig12}. The initial conditions of the disk model were successfully applied to explain high-angular resolution multi-wavelength observations of circumstellar disks, for example for HH30, CB26, and the Butterfly star \citepads[e.g.,][]{2003ApJ...588..373W,2008ApJ...674L.101W,2008A&A...478..779S,2009A&A...502..367S,2009A&A...505.1167S,2012A&A...543A..81M,2012A&A...546A...7L, 2013A&A...553A..69G}. Based on the disk density and opacity structure, the corresponding dust temperature distribution is calculated self-consistently with the continuum radiative transfer code MC3D \citepads{1999A&A...349..839W,2003ApJ...588..373W}, which is also used for post-processing the results of the MHD simulations.
The magnetic resistivity profile is calculated using the semi-analytical chemical model by \citet{dzy13}. In this work, we focus on the Ohmic diffusion term.
The structure of the paper is as follows: In the second section we describe our method and the physical model. The third section presents the results and the analysis of the simulations. Post-processing with radiative transfer and ALMA maps are presented in Sect. 4, followed by the discussion and the conclusion.  
%
%
%\begin{table*}
%\caption{Stellar and disk parameter. \label{tab:par}}
%\begin{tabular}{ll}
%star parameter & disk parameter \\
%\hline
%\hline
%0.95 $\rm L_\sun$ & $\rm \Sigma=$5.94 g $\rm cm^{-2}$ $\rm (r/100AU)$ \\
%$\rm T_*=4000K$ &    D2G ratio 0.01/0.0001                            \\
%$M_*=0.5 M_\sun$ &                                                    \\
%\hline
%\end{tabular}
%\end{table*}
%
%
\section{Methods and model}
We solved the non-ideal MHD equations using the PLUTO code \citep{mig07} with the orbital advection scheme FARGO MHD \citep{mig12}.
The equations are
\begin{equation}
\rm \frac{\partial \rho}{\partial t} + \nabla \cdot (\rho \vec{v}) = 0,
\label{eq:MDH_RHO}
\end{equation}
\begin{equation}
\rm \frac{\partial \rho \vec{v}}{\partial t} + \nabla \cdot \left [ \rho \vec{v} \vec{v}^T - \vec{B}\vec{B}^T \right ] + \rm \nabla P_t = - \rho \nabla \Phi,
\label{eq:MDH_MOM}
\end{equation}\
\begin{equation}
\rm \frac{\partial E}{\partial t} + \nabla \cdot \left [ (\rm E + P_t)\vec{v} - (\vec{v}\cdot\vec{B})\vec{B} \right ]  = - \rho \vec{v} \cdot \nabla \Phi -\nabla \cdot \left [ (\eta \cdot J) \times B \right ],\, and
\label{eq:MDH_EN}
\end{equation}
\begin{equation}
\rm \frac{\partial B}{\partial t} + \nabla \times (\vec{v} \times \vec{B}) = -\nabla \times(\eta \cdot J)
\label{eq:MDH_MAG}
\end{equation}
with the gas density $\rm \rho$, the velocity vector $\rm \vec{v}$, the magnetic field vector $\rm \vec{B}$, the total pressure $\rm P_t = P + 0.5B^2$, the gravitational potential $\rm \Phi$, the total energy $\rm E=\rho \epsilon + 0.5\rho \vec{v}^2 + 0.5 \vec{B}^2$  with the internal energy $\rho \epsilon = P/(\Gamma -1) $, the current density $\rm J= \nabla \times B$, and the Ohmic resistivity $\rm \eta$. We considered Ohmic diffusion. The importance of other magnetic diffusion terms is described in the discussion section.
%We use an adiabatic equation of state and the internal energy follows $\rho \epsilon = P/(\Gamma -1)$. 
We used a locally isothermal equation of state and fixed $\rm \Gamma =1.0001$. 
%for the isothermal runs and $\Gamma =1.42$ with a simple cooling function. 
The numerical MHD configuration was taken from \citet{flo12}. We use the constrained transport module which conserves $\nabla \cdot B = 0$ to machine precision \citep{mig07}. We used the HLLD Riemann solver \citep{miy05}. The Courant number was fixed to 0.3.

We continued radial and poloidal gradients of density, pressure, rotation velocity, and resistivity for radial and poloidal boundary conditions. For the radial and poloidal velocity we used a zero gradient with a linear damping of the normal component if it was pointing inward. In the poloidal direction we prevented the density from increasing in the ghost cells. This boundary condition allows for small inward accretion, which slightly reduces the total mass-loss rate.
\begin{figure}
\psfig{figure=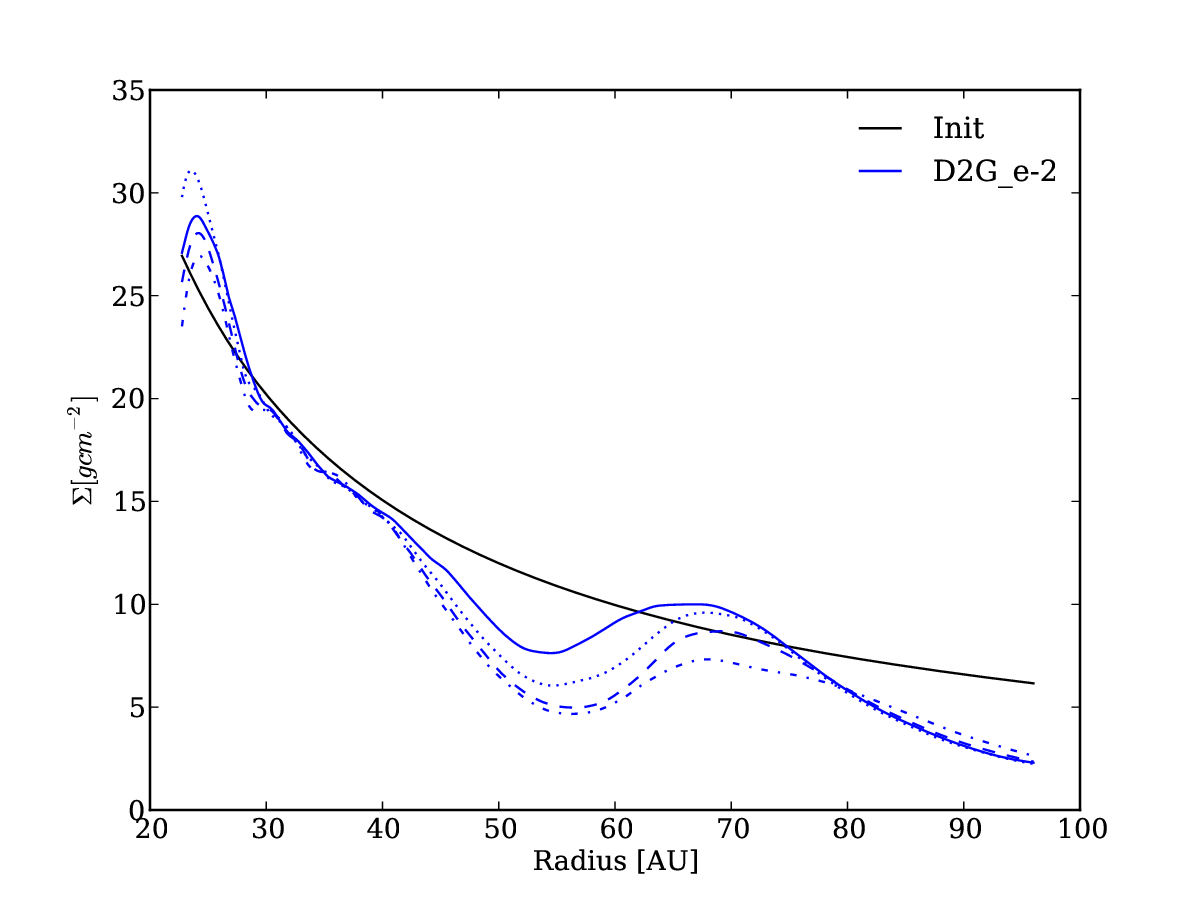,scale=0.40}
\psfig{figure=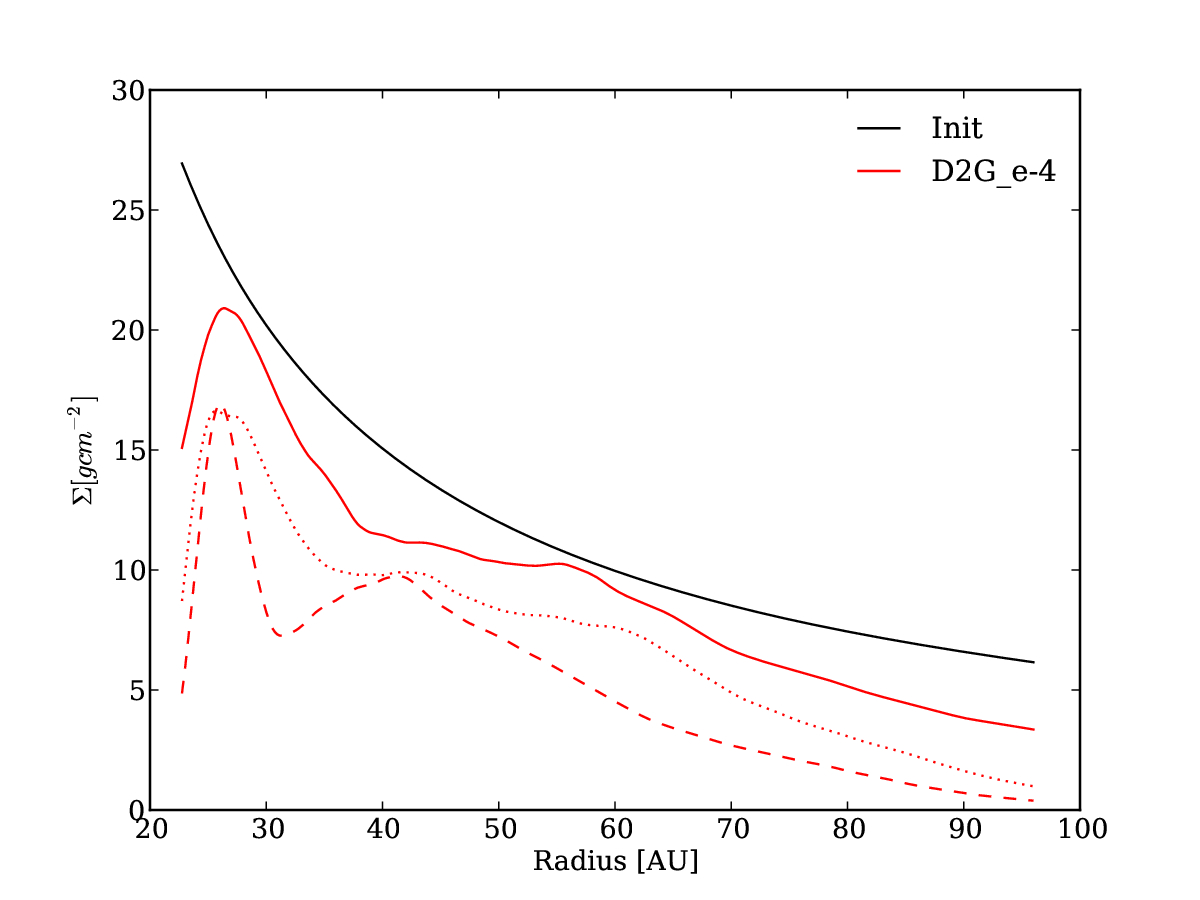,scale=0.40}
\caption{Radial surface density profiles, plotted in steps of 200 inner orbits, from initial ({\it black solid}) to final ({\it solid, dotted, dashed, and dashed-dotted}) for model \texttt{D2G\char`_e-2} ({\it blue}) and model \texttt{D2G\char`_e-4}({\it red}).}
\label{fig:SURF1D}
\end{figure}
\begin{figure}
%\vspace{-1cm}
\psfig{figure=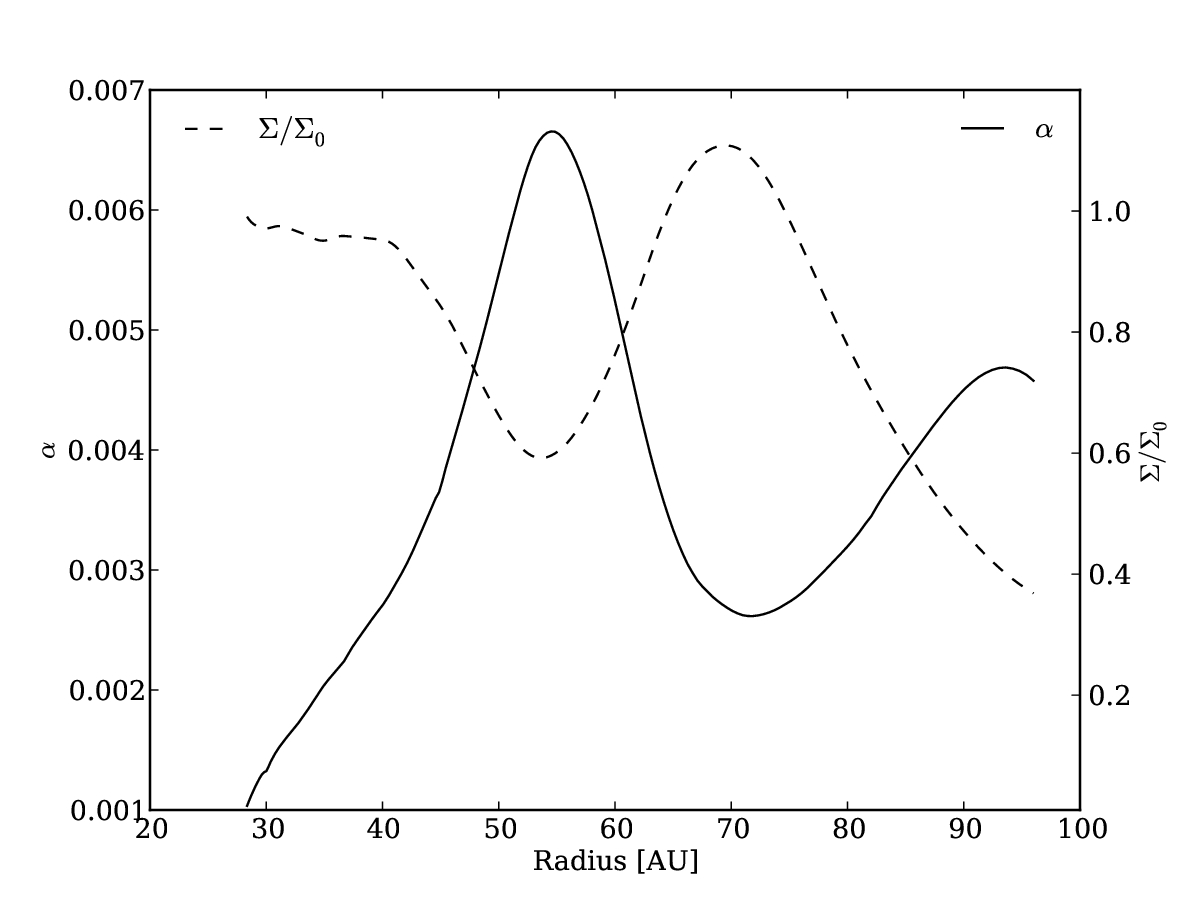,scale=0.40}
\psfig{figure=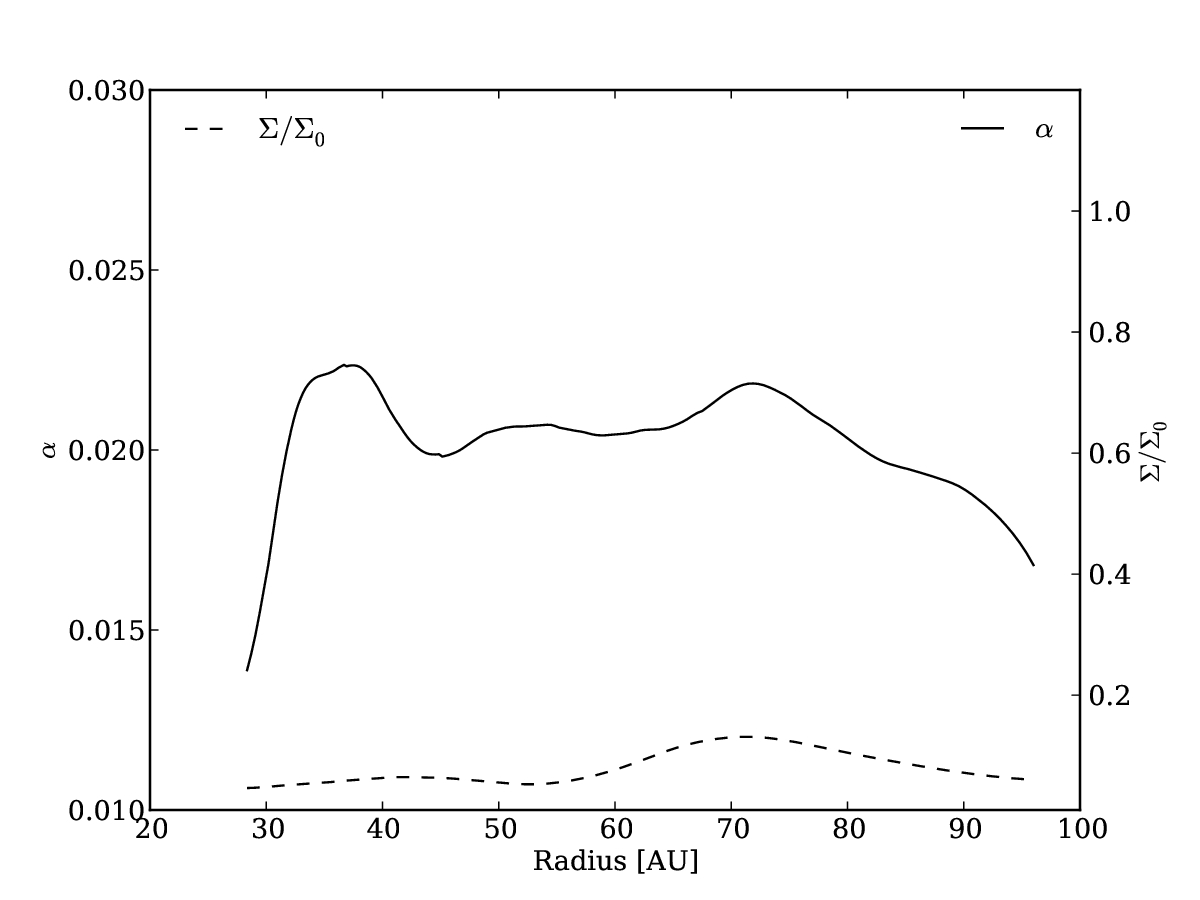,scale=0.40}
\caption{Time-averaged radial profile of the accretion stress $\rm \alpha$ ({\it solid line}) and the normalized surface density $\rm \Sigma/\Sigma_0$ ({\it dashed line}) for models \texttt{D2G\char`_e-2} ({\it top}) and \texttt{D2G\char`_e-4}({\it bottom}).}
\label{fig:alpha}
\end{figure}
\subsection{Disk model}
The initial conditions and parameters of the stellar system are summarized in Table~\ref{tab:mod} and Fig.~\ref{fig:INIT}. The initial gas surface density profile $\Sigma_0$ follows
\begin{equation}
\rm \Sigma_0=5.94 g\, cm^{-2} \left( \frac{100AU}{R} \right),
\end{equation}
with the cylindrical radius R.
%Initial conditions are plotted in Fig.~\ref{fig:INIT}.
Temperature and density profiles are in fully hydrostatic equilibrium. The radial domain spans from 20 to 100 AU. We used a logarithmically increasing grid in the radial direction. The vertical extent was $\Delta \theta=0.72$ radian. The azimuthal domain spans a full $360^\circ$. The local scale height was well resolved with around 20 grid cells in each direction given by the H/R value of about 0.1. A resolution study is presented in Appendix B. The initial vertical magnetic field followed $\propto \rm R^{-1}$. We defined the initial magnetic vector potential $\rm A_\phi=A_0$ and the magnetic field followed $\rm \vec{B}_{init} = \nabla \times A$. The strength of the magnetic field was set with $\rm A_0$ by the condition $\rm \beta_{min} > 2.0$ in the full domain with the plasma beta $\beta = 2 P/(\vec{B}^2)$. The initial magnetic field strength was about 1 mGauss at 40 AU, which is close to the strength of net vertical field predicted from theoretical models \citep{oku14}. The initial resistivity profile was calculated using the chemistry method by \citet{dzy13}. 
The abundances of the main ion, electrons, and charged dust were obtained assuming that the chemistry quickly reaches the charge equilibrium in the presence of small fractal dust \citep{oku09}. \label{sec:chemistry}
 For the dust chemistry, we considered representative fractal dust aggregates, consisting of 400 monomers of size $\rm 0.1 \mu$m\footnote{We note that for 2D fractal dust, the size of the monomer is more important than the size of the aggregate to determine the ionization degree \citep{oku09}. We chose the monomer size by adopting the values from \citet{war07}. The effect of the monomer size on the ionization profile remains small: changing the monomer size by two orders of magnitude would increase the extent of the dead zone by a factor of 2 \citep{dzy13}.}. Such fractal dust particles dominate in the ability to remove free electrons from the surrounding gas \citep{oku09}.
%with the internal density of $\rm 1.4\rm g\, cm^{-3}$, what corresponds to 'icy' dust, typical for outer disks. 
Metals are frozen out, and the representative ion is $\rm HCO^+$ \citep{hen13,dut14}. 
We adopted the X-ray ionization rate from \citet{bai09} ( case of 3 keV).
The cosmic-ray ionization rate outside of the disk was $5\times 10^{-18} \rm s^{-1}$. Ionization due to radio-nuclide was $\rm 7\times{}10^{-19} \rm s^{-1} \, (D2G/10^{-2})$, where $\rm D2G $ is dust--to--gas mass ratio. We used two different dust--to--gas mass ratios, $10^{-2}$ and $10^{-4}$, corresponding to models \texttt{D2G\char`_e-2} and \texttt{D2G\char`_e-4}. The two profiles of the magnetic Reynolds number $\rm Re_m=c_s H/\eta$ are plotted in Fig.~\ref{fig:INIT} for the fiducial ({\it top right}) and the dust-depleted ({\it bottom right}) disk model. 
\begin{table*}
\caption{Top: Model name, resolution, domain size, dust--to--gas mass ratio, run time, averaged accretion stress, and averaged azimuthal surface density variations. Bottom: Disk and stellar parameters. \label{tab:mod}}
\begin{tabular}{lllllll}
Model name & $\rm N_r x N_\theta x N_\phi$ & $\rm \Delta r\, [AU] : \Delta \theta\, [rad] : \Delta \phi\, [rad]$ & D2G & Inner orbits & $\rm <\alpha>$ & $\rm <\Sigma'>$\\  \hline
\hline
\texttt{D2G\char`_e-4}  & $256$x$128$x$512$ & 20-100 : 0.72 : $2\pi$ & $10^{-4}$ &800 & 0.013 & 0.04\\ 
\texttt{D2G\char`_e-2}  & $256$x$128$x$512$ & 20-100 : 0.72 : $2\pi$ & $10^{-2}$ &1045 & 0.003 & 0.09\\ 
& & & & & &\\
Disk parameter &  & Stellar parameters & & & &\\
\hline
\hline
$\rm M_{total} \cong 0.085 M_*$ & $\rm \Sigma_0=$5.94 g $\rm cm^{-2}$ $\rm \frac{100AU}{R}$ & $\rm T_*=4000K$ & 0.95 $\rm L_\sun$  & $\rm M_*=0.5 M_\sun$ & spectral type $\sim$K  & \\
%0.95 $\rm L_\sun$ & $\rm \Sigma=$5.94 g $\rm cm^{-2}$ $\rm (r/100AU)$ & & & & & \\
%$\rm T_*=4000K$ &    D2G ratio 0.01/0.0001 & & & & &                            \\
%$M_*=0.5 M_\sun$ &        & & & & &                                            \\

\end{tabular}
\end{table*}
\section{Results}
In Table~\ref{tab:mod} we summarize the setup and model parameters. 
Because of the strong linear MRI phase, which triggers a strong mass outflow, we reset the density to its initial profile after 200 inner orbits. We average the time outputs during the simulation run time after the restart. More details about the linear phase can be found in the appendix.

\begin{figure*}
\hspace{-1cm}
\psfig{figure=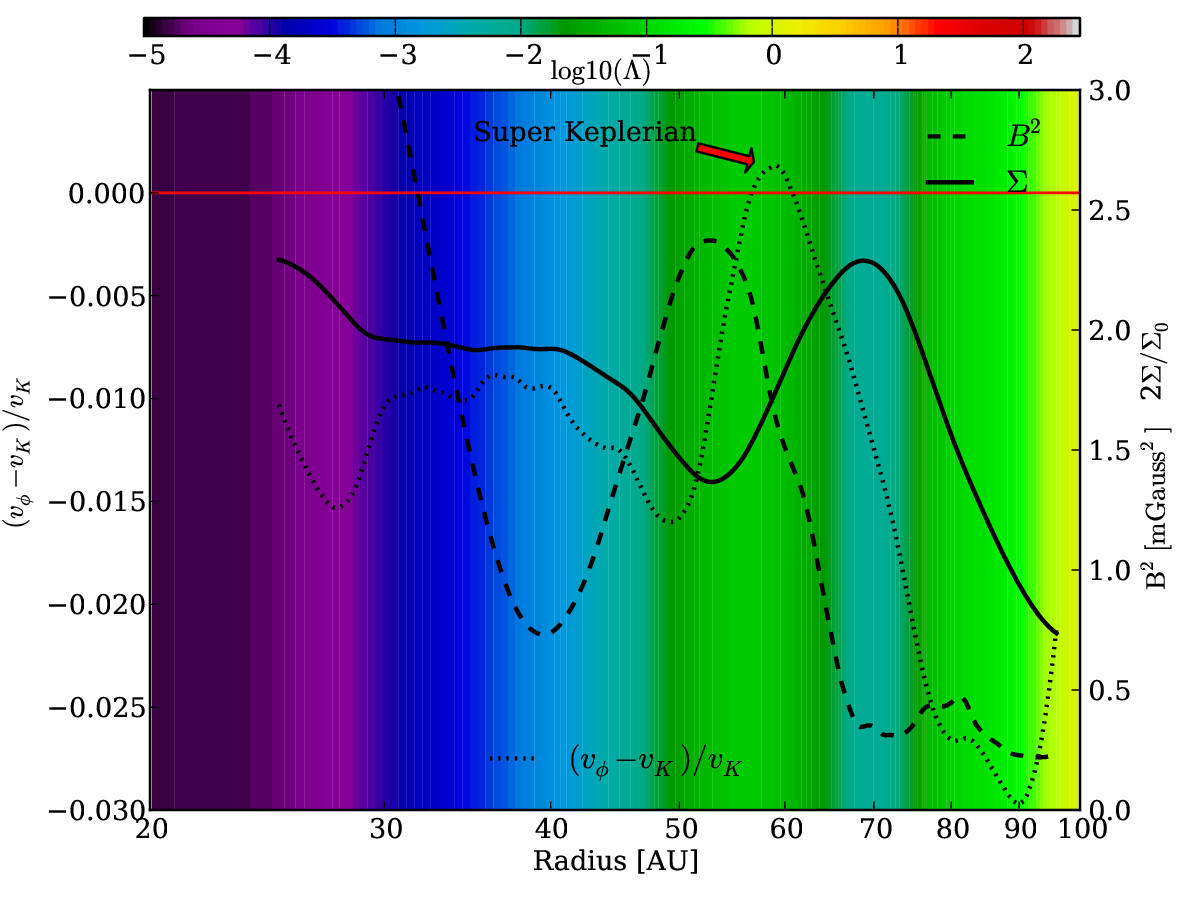,scale=0.47}
\psfig{figure=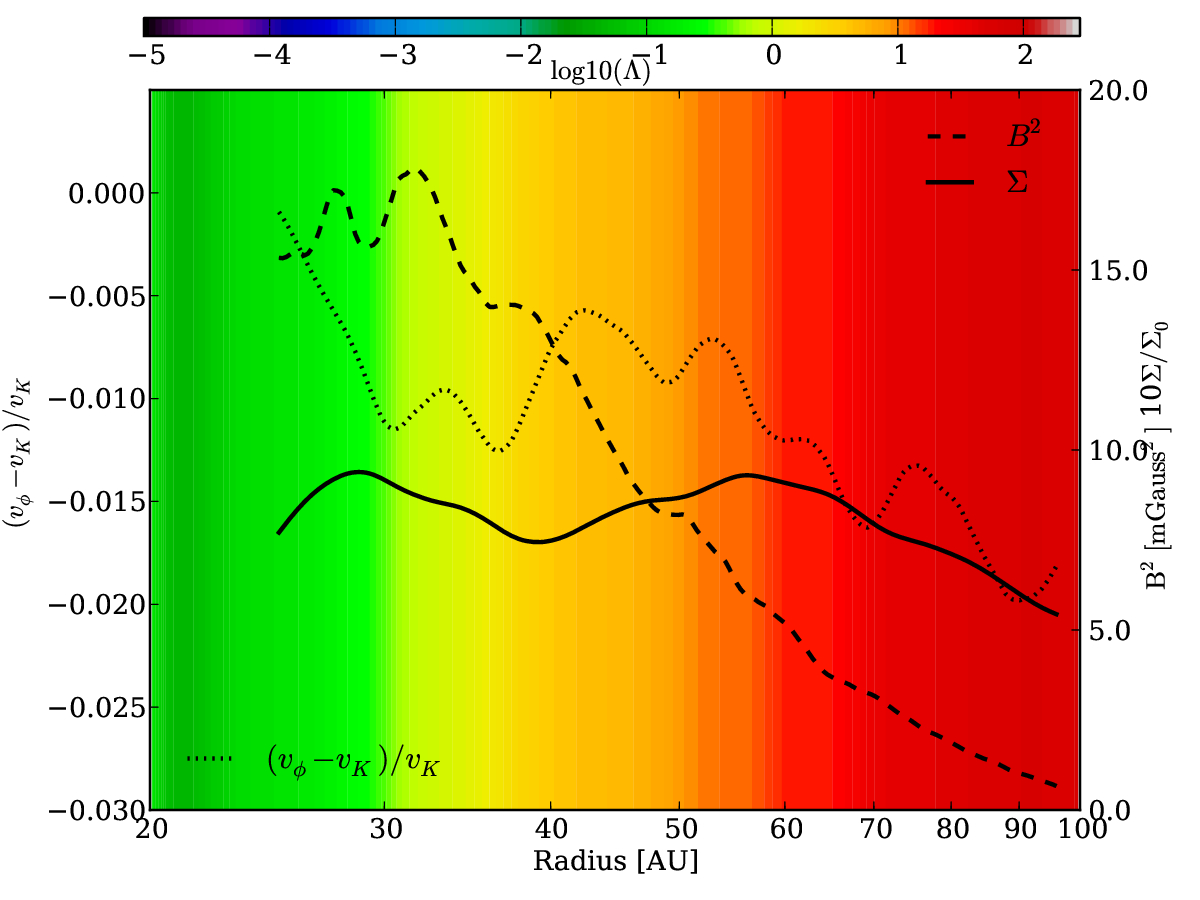,scale=0.47}
\caption{Radial profiles of space and time averaged surface density ({\it solid line}), magnetic pressure ({\it dashed line}) and deviation of Keplerian rotation ({\it dotted line}) for model \texttt{D2G\char`_e-2} ({\it left}) and model \texttt{D2G\char`_e-4} ({\it right}). The background contour color shows the space and time averaged value of the Elsasser number. The red solid line and annotation show the region of super Keplerian rotating gas. At this location, large particles are getting trapped.}
\label{fig:EL}
\end{figure*}

The two models quickly established a turbulent state, driven by the MRI. 
The time-averaged accretion stress was calculated with
\begin{equation}
\rm \alpha = \left\langle \frac{ \int \Bigg(
\frac{\rho v'_{\phi}v'_{r}}{c_s^2} -
\frac{B_{\phi}B_{r}}{c_s^2}\Bigg)dV} {\int \rho dV} \right\rangle.
\label{eq:ALPHA}
\end{equation}
The turbulent radial and azimuthal velocities were calculated by subtracting the azimuthally averaged velocities from the total one, $\rm v_{r,\phi}'=v_{r,\phi}- \overline{v_{r,\phi}}_{\phi}$. This averaging is important because the temperature and consequently, the sound speed $\rm c_s$ depend on the radius and height. Along the azimuthal direction, $\rm c_s$ is constant and the MRI generated turbulence is self-similar. Model \texttt{D2G\char`_e-4} shows a strong turbulent state with a time-averaged value of 0.013, while in the fiducial model \texttt{D2G\char`_e-2}, the averaged accretion stress is reduced to 0.003 as a result of the higher resistivity. The variations of the surface density were measured by 
\begin{equation}
<\Sigma'>=< \overline{ \left| \Sigma-\overline{\Sigma_{\phi}} \right |}/ \overline{\Sigma_{\phi}}>,
\end{equation} 
with $\rm \overline{\Sigma_{\phi}}$ being the surface density averaged over azimuth and $<>$ presenting the space and time average over the full domain and run time. The fluctuations along azimuth are around 4 percent in model \texttt{D2G\char`_e-2}, while they increase to 9 percent in model \texttt{D2G\char`_e-4}. Both values are still low and difficult to resolve with current telescope facilities; see more in Sect. 4. The total surface density over radius evolves distinctly for both models. In Fig.~\ref{fig:SURF1D} we show the surface density over radius, plotted in steps of 200 inner orbits. 
In model \texttt{D2G\char`_e-4} the surface density quickly decreases with time as a result of the strong turbulence. In contrast, model \texttt{D2G\char`_e-2} shows a gap and jump structure that develop between 50 and 80 AU. The emerging feature is present during the whole simulation time. 
Time-averaged radial profiles of the accretion stress and the normalized surface density $\rm \Sigma/\Sigma_0$ are plotted in Fig.~\ref{fig:alpha}. The gap structure in model \texttt{D2G\char`_e-2} is highly turbulent, while in the jump the turbulence drops by a factor of 2. Radially inward, the higher resistivity starts to dampen the MRI and the accretion stress is reduced. Model \texttt{D2G\char`_e-4} shows a much stronger accretion stress, nearly constant over radius. Because of the strong turbulence, the surface density in model \texttt{D2G\char`_e-4} is quickly reduced. 

We summarize that we observed the formation of a distinct gap and jump structure in the surface density for our fiducial model with a dust--to--gas mass ratio of $10^{-2}$. This structure was anti--correlated with the accretion stress. The gap shows a highly turbulent state, while the jump is less turbulent. The model with a reduced dust amount shows a fully turbulent disk. Here, the strong turbulence leads to a fast removal of the disk. 
%
%
%
%
%\begin{figure}
%\psfig{figure=PS/Zfl2.jpg,scale=0.40}
%\caption{Azimuthal and vertical averaged radial profiles of $\partial R v_\phi/\partial R$ (solid black line), the turbulent magnetic field components (red) and the accretion stress components (blue) for model \texttt{D2G\char`_e-2}.}
%\label{fig:ZFL2}
%\end{figure}
%
%
\subsection{Strong zonal-flow structure at the dead-zone outer edge}
In this section we investigate the stability of the gap and ring structure by regarding the MRI activity at the dead-zone edge. 
In Fig.~\ref{fig:EL} we plot time- (300-500 inner orbits) and space- ($\rm \pm 0.5 H$) averaged radial profiles of magnetic pressure $\rm B^2/2$, and the normalized surface density  $\rm 2\Sigma/\Sigma_0$ and deviation of the Keplerian velocity $\rm (v_\phi-v_k)/v_k$. The Elsasser number $\rm \Lambda_z=v_a^2/(\eta \Omega)$ for the same time and space average is plotted as background contour color. The structure that emerges in the fiducial model \texttt{D2G\char`_e-2}, Fig.~\ref{fig:EL} ({\it left}), shows very similar characteristics as the classical zonal flow found in local-box simulations \citep{joh09,dit13} and global simulations \citep{dzy10,uri11}. There is a clear anti--correlation between the magnetic pressure and the surface density profile, also found in local box simulations, compare Fig.~7 by \citet{joh09}. The rise in surface density is so strong that the rotation velocity becomes super-Keplerian (see red solid line and annotation). This will stop and even reverse the radial drift of larger particles. At this location, we would expect a high concentration of solid material. Another important feature is the correlation between the Elsasser number with the radial profiles of surface density and magnetic pressure. Inside the gap, the Elsasser number peaks at 0.1, which enables MRI activity \citep{tur07,oku11,flo12}. Inside the bump, the Elsasser number is below 0.01, which is clearly not sufficient to enable the MRI. This indicates that the structure is self-sustained because the active part will continue to clear the gas out of the gap. However, the jump in surface density is stable against the MRI. Here, the RWI will eventually grow, while it will try to flatten the jump in surface density. 
In model \texttt{D2G\char`_e-2} we observed two cycles of a large vortex formation that lasted for roughly 40 local orbits. The regions around the jump in surface density are MRI turbulent and re-establish the bump, so that the cycle continues (see next section). 

\begin{figure*}[t]
\hspace{-0.5cm}
\psfig{figure=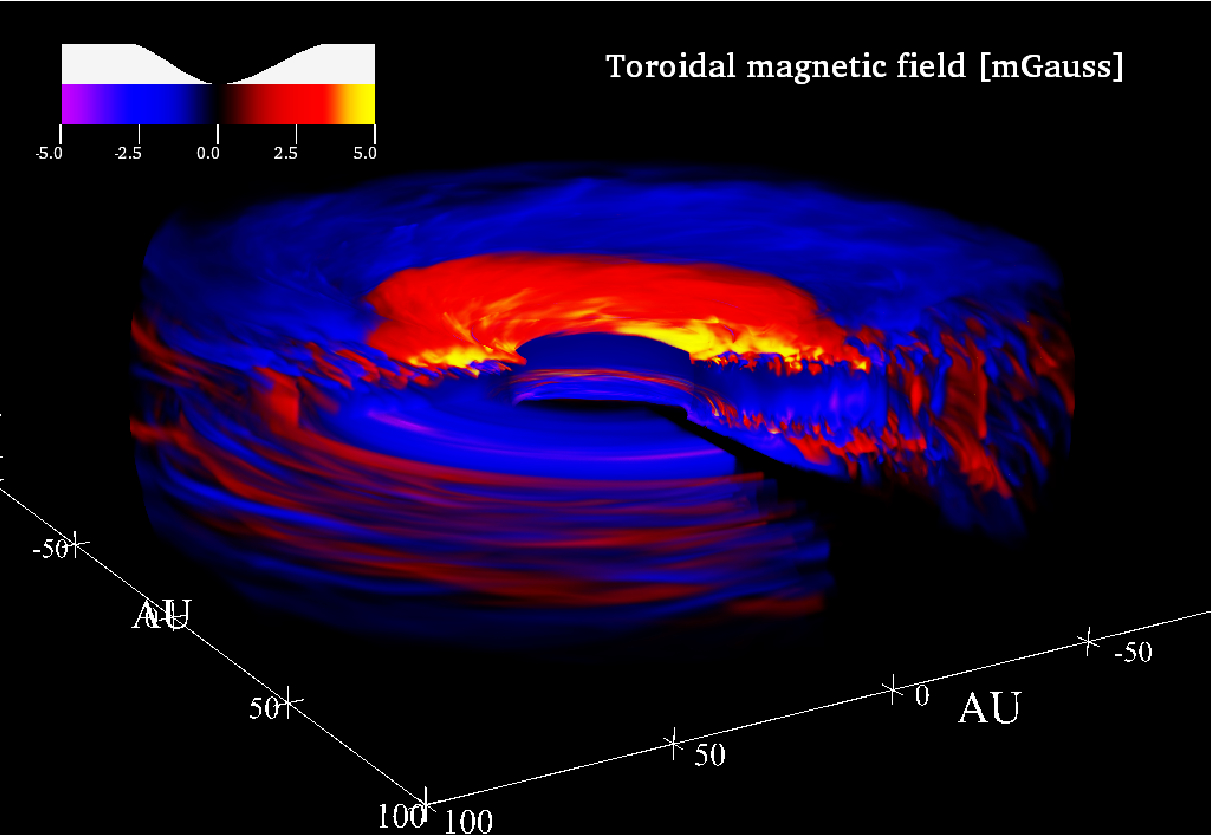,scale=0.47}
\psfig{figure=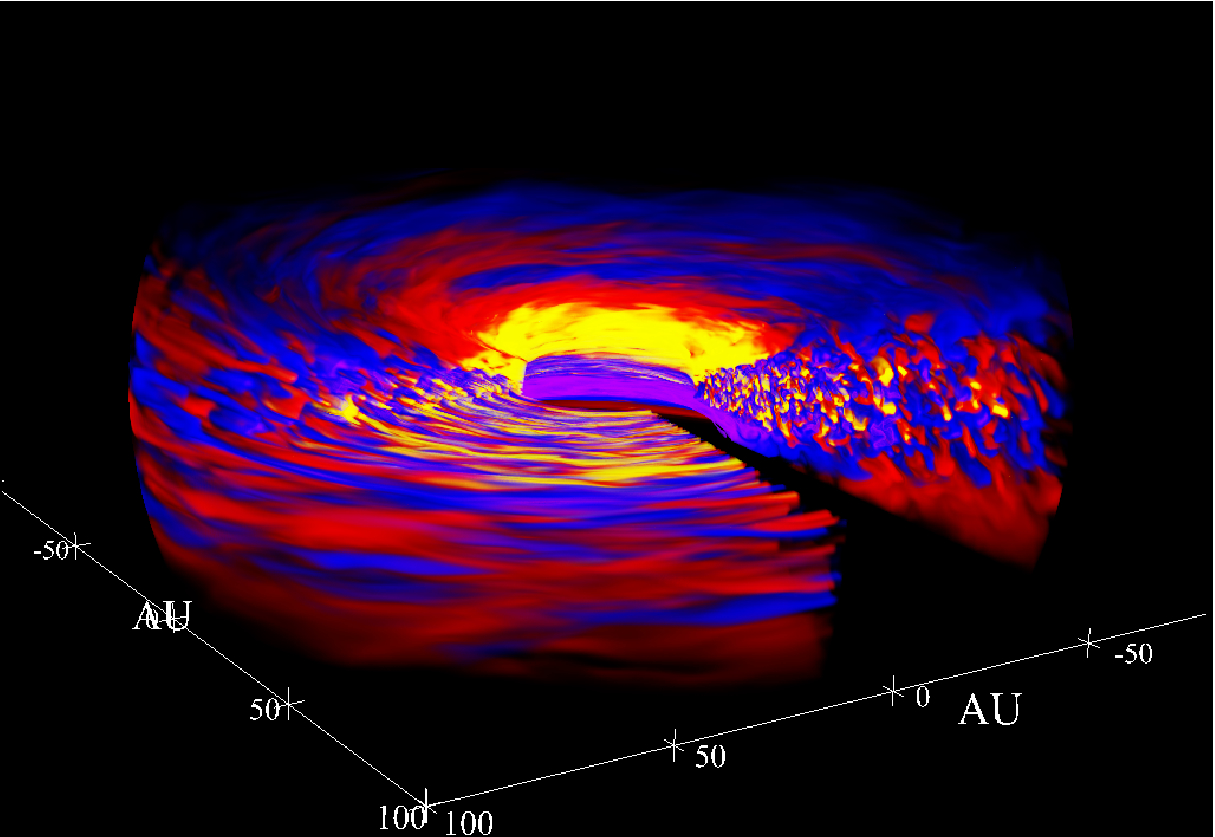,scale=0.47}
\caption{3D plot of the toroidal magnetic field of models \texttt{D2G\char`_e-2} ({\it left}) and \texttt{D2G\char`_e-4} ({\it right}).}
\label{fig:BPHR}
\end{figure*}

The magnetic fields consist mainly of the toroidal component. In both models, they are dominant in the whole disk with a strength of several mGauss.
%Fig.~\ref{fig:ZFL2} shows for the same time and space averaged radial profiles of $\partial R v_\phi/\partial R$ ({\it black solid line}), the turbulent magnetic components ({\it red lines}) and the accretion stress components ({\it blue lines}) for model \texttt{D2G\char`_e-2}. As expected we observe a peak in the gradient of the rotational velocity inside the gap of the zonal flow. 
%In this gap we observe a maximum of turbulent activity mainly to strong turbulent toroidal fields of the order of mGauss strength ({\it red dashed line}). 
A full 3D snapshot of the toroidal magnetic field is plotted in Fig.~\ref{fig:BPHR} for model \texttt{D2G\char`_e-2} ({\it left}) and model \texttt{D2G\char`_e-4} ({\it right}) after 410 inner orbits. 
Model \texttt{D2G\char`_e-4} shows a fully evolved turbulence with small-scale structures in the whole disk. The fiducial model \texttt{D2G\char`_e-2} shows a smooth field configuration due to the strong resistivity at the inner part, especially inside 50 AU. Both figures show that there is magnetic field accumulation inside the gap and the dead zone. For model \texttt{D2G\char`_e-4}, the Elsasser numbers is well above 0.1 and the MRI operates in the whole disk. Here, we do not observe any gap or jump structure. 
%But apart from the large scale structures like a zonal flow, particles could be also concentrated in local pressure perturbations \citep{dit13}. 
%
%
%
%
%
%
%\begin{figure*}[t]
%\psfig{figure=PS/LR_410_BP2.jpg,scale=0.85}
%\caption{3D snapshot of toroidal magnetic field in model \texttt{D2G\char`_e-4}.}
%\label{fig:BPLR}
%\end{figure*}
%
%
% TO READ HERE
%
\subsection{Axisymmetric and non-axisymmetric structures}
\begin{figure*}[t]
\hspace{-0.5cm}
\begin{minipage}{0.4\textwidth}
\psfig{figure=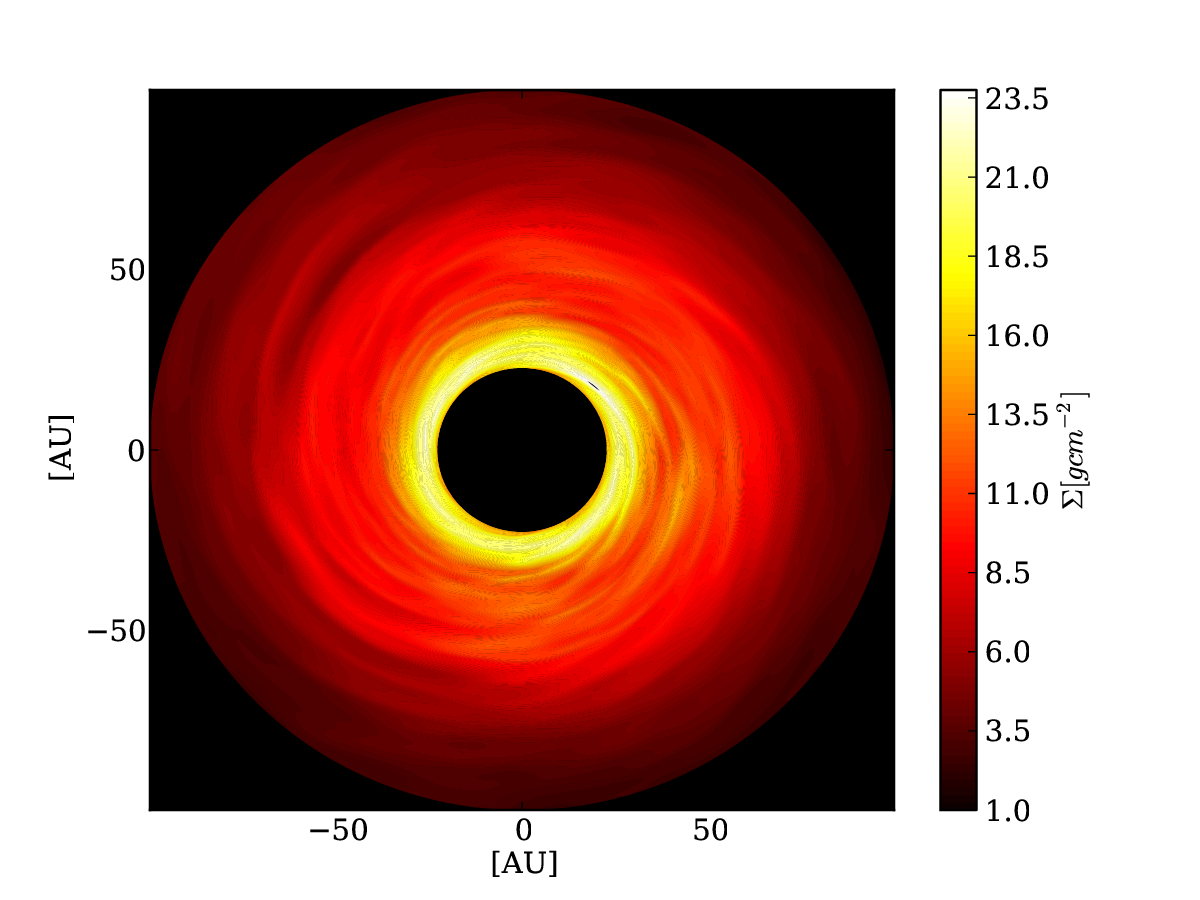,scale=0.5}
\psfig{figure=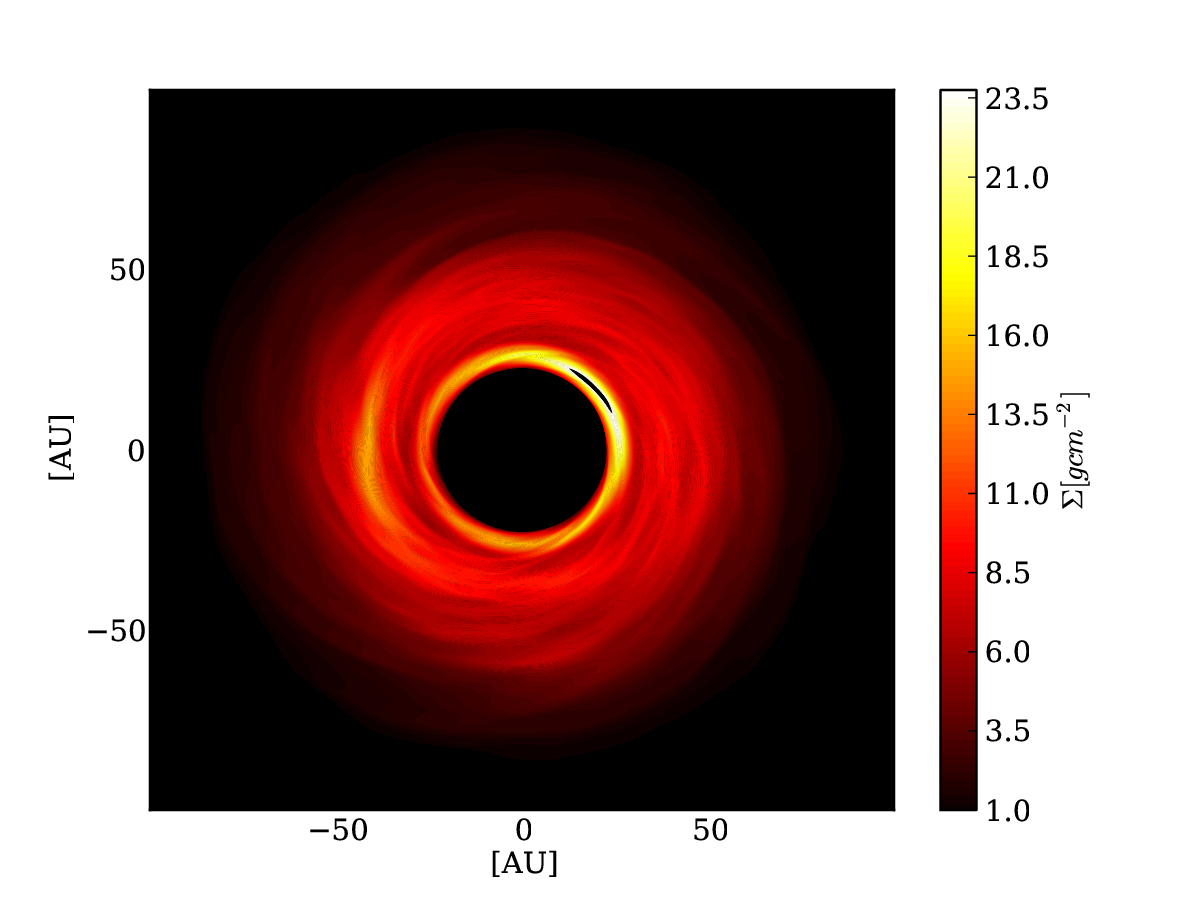,scale=0.5}
\end{minipage}
\hspace{2.0cm}
\begin{minipage}{0.4\textwidth}
\psfig{figure=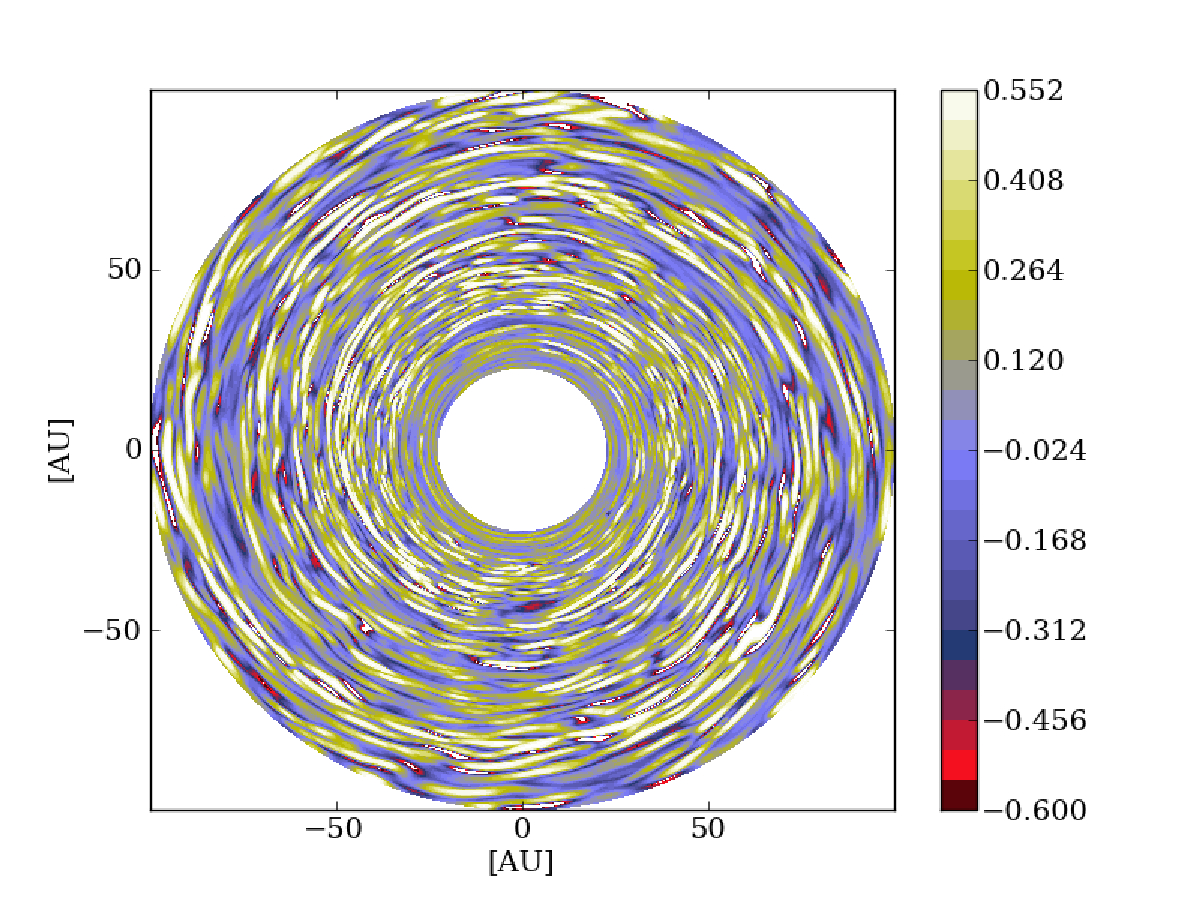,scale=0.5}
\psfig{figure=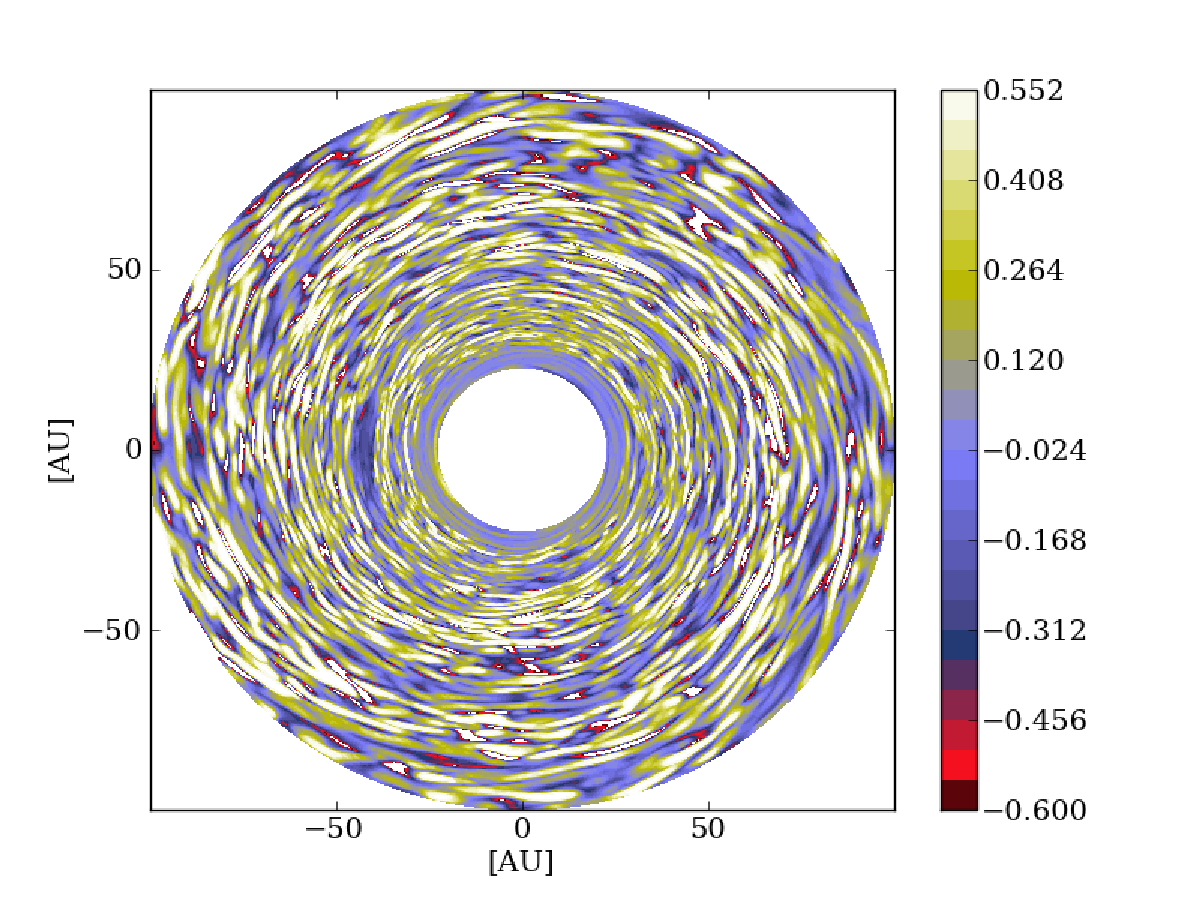,scale=0.5}
\end{minipage}
\caption{Surface density maps ({\it left}) for two different time outputs (400 and 800 inner orbits) for model \texttt{D2G\char`_e-4} with the corresponding vorticity maps ({\it right}).}
\label{fig:SURFLR}
\end{figure*}

\begin{figure*}[t]
\hspace{-0.5cm}
\begin{minipage}{0.4\textwidth}
\psfig{figure=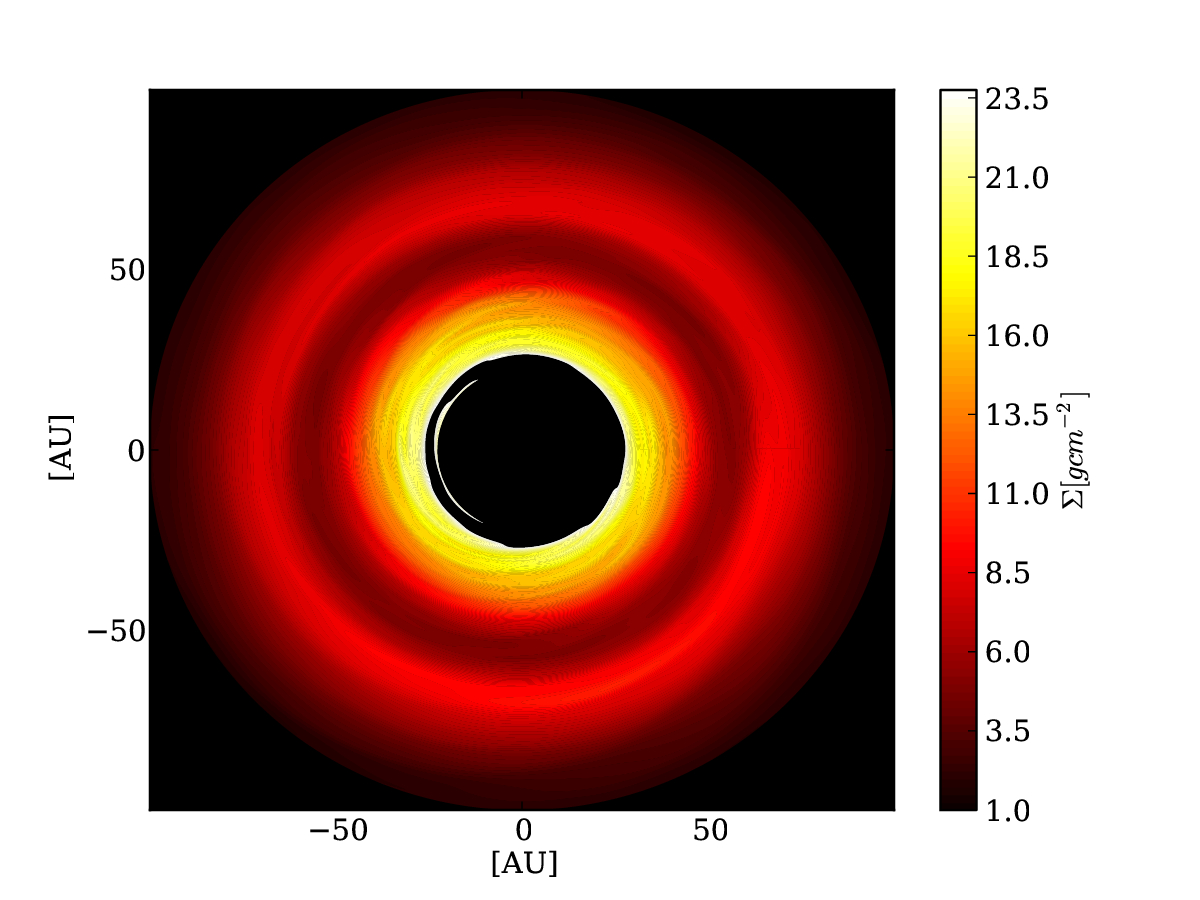,scale=0.5}
\psfig{figure=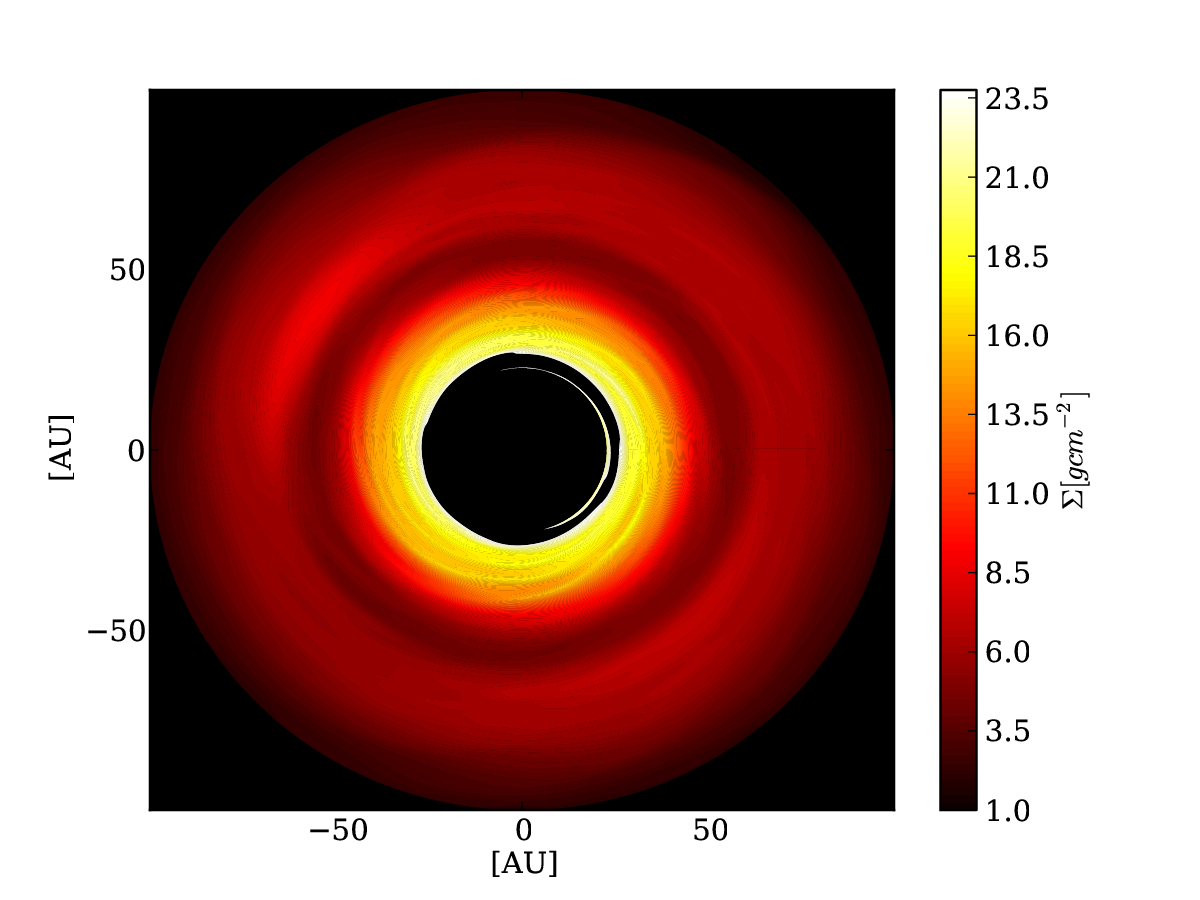,scale=0.5}
\end{minipage}
\hspace{2.0cm}
\begin{minipage}{0.4\textwidth}
\psfig{figure=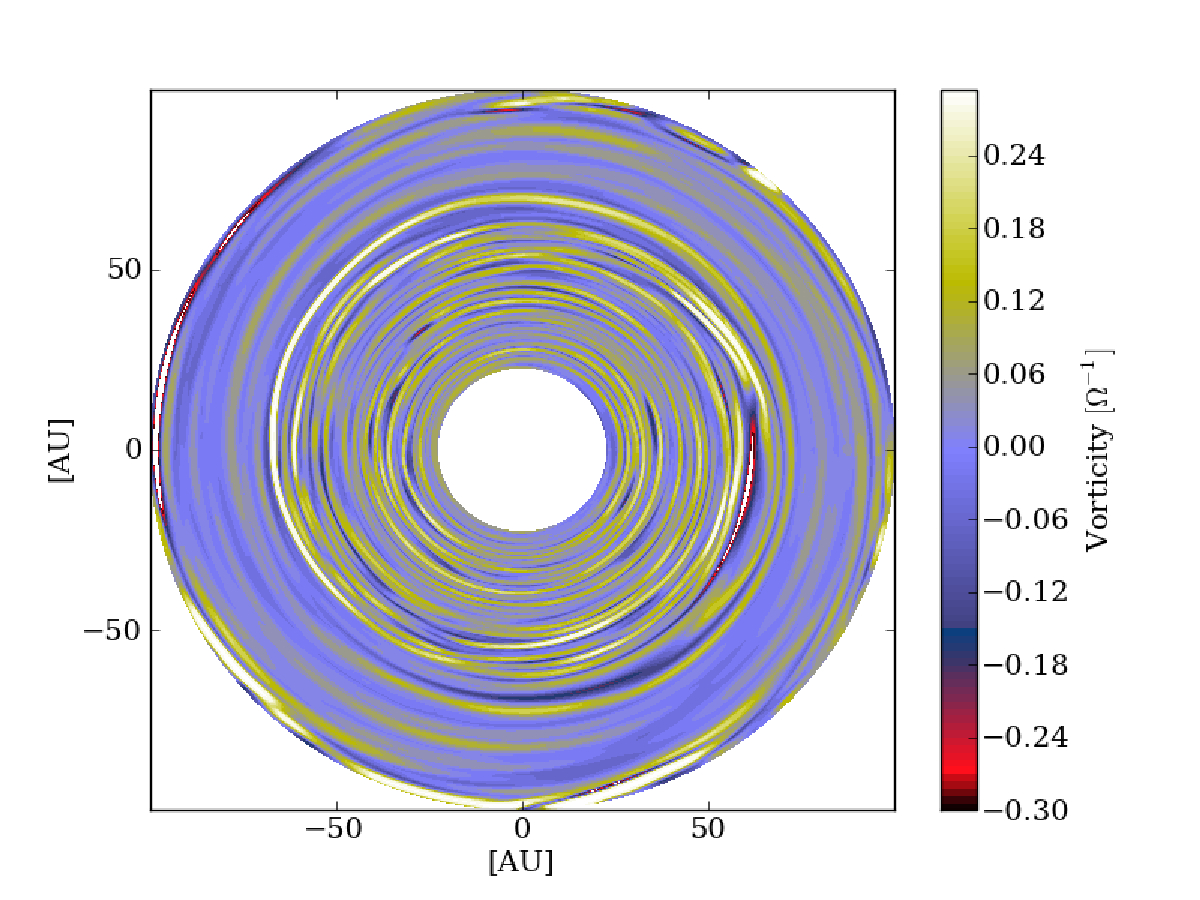,scale=0.5}
\psfig{figure=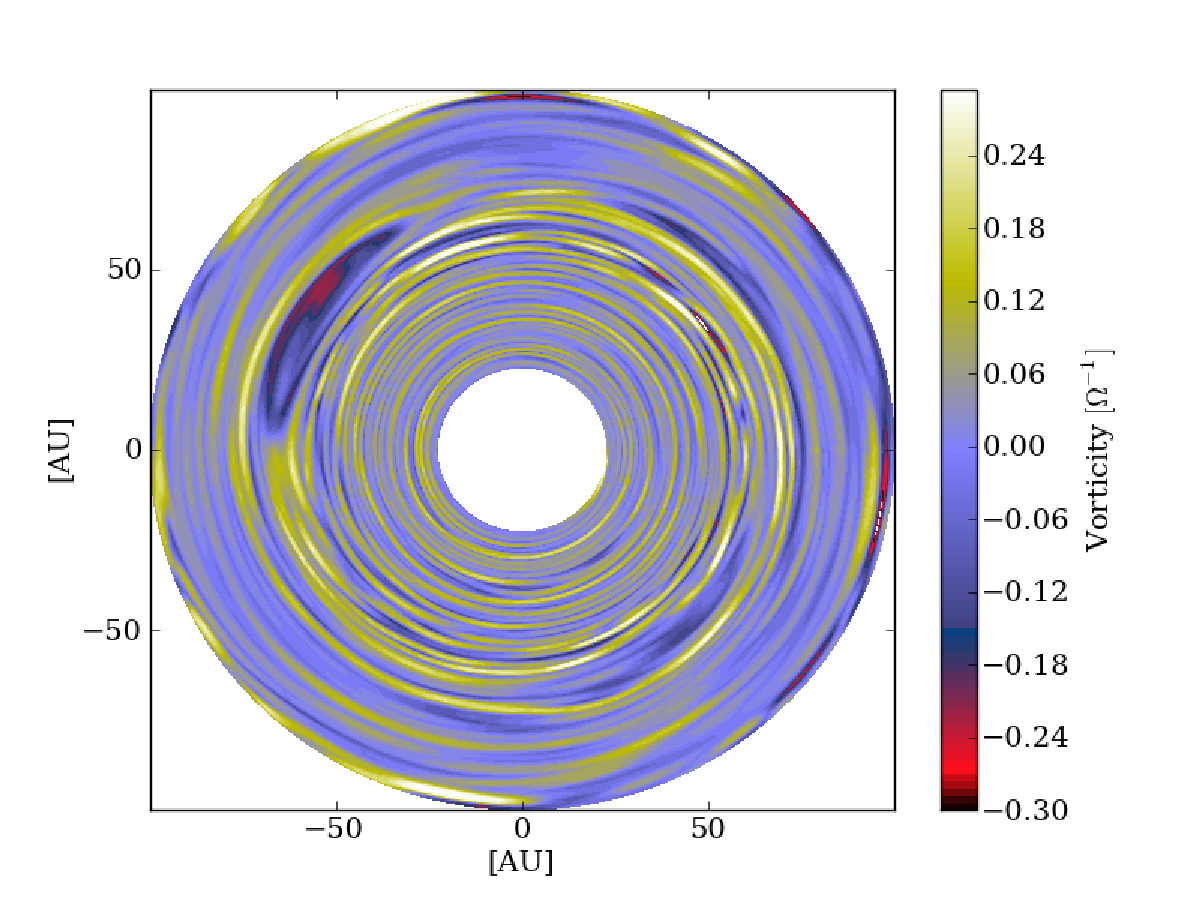,scale=0.5}
\end{minipage}
\caption{Surface density maps ({\it left}) for two different time outputs (800 and 1045 inner orbits) for model \texttt{D2G\char`_e-2} with the corresponding vorticity maps ({\it right}). }
\label{fig:SURFHR}
\end{figure*}

In the fully turbulent model \texttt{D2G\char`_e-4}, the surface density has clear non-axisymmetric structures, which is shown in the surface density maps in Fig.~\ref{fig:SURFLR} after 400 ({\it top}) and 800 ({\it bottom}) inner orbits. 
As shown in Table~\ref{tab:mod}, the strength of these surface density fluctuations are around 10 \%, which is not high enough to be resolved by current telescopes; see Sect. 4. 
%
%  TO READ HERE
%
Here, local enhancements in the surface density with a small spatial extent can reach up to 50 \%, e.g. in Fig.~\ref{fig:SURFLR} ({\it bottom left} close to 40 AU). The corresponding maps of the vorticity $\rm (\nabla \times v)_z$ at the midplane, Fig.~\ref{fig:SURFLR} ({\it bottom right}), shows that these large-scale structures correlate with a minimum in vorticity. These structures could potentially concentrate particles, similar to concentrations of particles in fully ionized MRI turbulence \citep{dit13}.\\
% CALCULATE MEAN AZIMUTAHL SURFACE DENSITY VARIATION
Surface density and vorticity maps of model \texttt{D2G\char`_e-2} are presented in Fig.~\ref{fig:SURFHR} after 800 and 1045 inner orbits. Both outputs show the clear gap and following jump in surface density, visible between 50 and 80 AU. The structures are mostly axisymmetric, with small azimuthal fluctuations of about 4 \%. In this model we observe the formation of a large vortex inside the ring structure at 60 AU (see Fig. \ref{fig:SURFHR} {\it bottom}), with a surface density enhancement and a corresponding low-vorticity region. The radial extent of the vortex is around 2 scale heights, which corresponds to a radial extent of 10 AU at this location. The azimuthal extent is around 10 H, corresponding to an extent of 50 AU. We tracked two vortices in model \texttt{D2G\char`_e-2} with similar strength over the full simulation. The first has a lifetime of around 40 local orbits at 60 AU, starting after 350 inner orbits and lasting for 200 inner orbits. The second cycle starts at around 850 and is still present at the final output; see Fig.~\ref{fig:SURFHR}.
The vorticity amplitude inside the vortices is around -0.3 $\rm [\Omega^{-1}]$, which matches those generated by the RWI \citep{meh12,meh13}. 

We summarize that both models developed surface density enhancements. In the fully turbulent model \texttt{D2G\char`_e-4} the enhancements are stronger, but with a smaller spatial extent than for model \texttt{D2G\char`_e-2}. In the fiducial model \texttt{D2G\char`_e-2} we observe a gap followed by a bump in surface density. Inside the bump, vortices are formed by the RWI with a lifetime of around 40 local orbits.
%Independent of their generation process, vortices are able to concentrate particles \citep{meh12,meh13}. Especially larger particles with a coupling time (Eq. 8) close to the rotational time, could be quickly concentrated (compare Fig. 4 of \citep{meh12}). 
% as well as inside the ring structure of model \texttt{D2G\char`_e-2} in Fig.~\ref{fig:SURFHR} ({\it bottom}). Such overdensities are created by vortices. We calculated the vorticity component $(\nabla \times v)_z$ at the midplane for both models. Fig.~\ref{fig:VORTI} shows 2D maps of $(\nabla \times v)_z$ in the $r-\phi$ plane. 
% PLot vortices maps for same output
%
%
%
%Model \texttt{D2G\char`_e-2} shows the generation of vortices inside the ring structure between 60 and 80 AU (see Fig.~\ref{fig:VORTI}, right). At 80 AU after 200 inner orbits (25 %local orbits), there is a large vortex formed (see Fig.~\ref{fig:VORTI}, top right). 50 local orbits later the vortex has disappeared (Fig.~\ref{fig:VORTI}, middle right). The latest output%, 200 inner orbits later shows a new formed vortex between 60 and 70 AU which is correlated with the overdensity in Fig.~\ref{fig:SURF}, bottom right. 
%
%Interesting is that also the fully turbulent model \texttt{D2G\char`_e-4} shows the generation and destruction of vortices.  Fig.~\ref{fig:VORTI}, left, shows vortices embedded inside the turbulent structures generated by the MRI. Strength and size are of similar values as in model \texttt{D2G\char`_e-2}. 
%
%

\section{Radiative transfer and simulated ALMA maps}
\label{sec:rt}

The fiducial model showed the generation of a gap and bump in surface density, visible as ring structures located at the outer edge of the dead zone. 
The decrease and following enhancement of the surface density emerges thanks to a zonal-flow-analog structure, which is located at the transition between the MRI dead and active zone. 
The model with a reduced dust amount showed fully evolved turbulence. In this chapter we investigate and compare these two model configurations when observed with ALMA. 
\subsection{Radiative transfer and ALMA simulations}
\label{sec:RT}
\paragraph{The radiative transfer (RT)} calculations of the temperature distribution of the dust phase and the corresponding thermal re-emission maps were performed with the Monte Carlo-based 3D continuum RT code MC3D \citepads{1999A&A...349..839W,2003CoPhC.150...99W}. The grid structure in the RT simulations was identical to those of the MHD simulations. The radiation source was a T Tauri star in black-body approximation. It is characterized by its luminosity ($0.95\, \rm L_\odot$) and effective surface temperature ($4000\, \rm K$).\par

We focus on the continuum thermal dust emission of disks in face-on orientation. Small dust particles ($< 0.25\, \rm \mu m$) couple to the gas \citepads{2004A&A...425L...9P}. Therefore, we derived the spatial dust distribution from the gas density distribution by applying the global dust--to--gas mass ratios listed in Table~\ref{tab:mod}. The dust is composed of $62.5\, \%$ silicate and $37.5\, \%$ graphite (optical data by \citealtads{2001ApJ...548..296W}) with a density of the dust material of $\rho_\text{dust} = 2.7\, \rm g\,cm^{-3}$. The grains are spherical with a grain size distribution following the power-law $\text{d}\, n(a_\text{dust})/\text{d}\, a_\text{dust} \propto {a_\text{dust}}^{-q}$ for a grain size exponent $q =3.5$ within the range $a_\text{dust} \in \left[0.005\, \mu \textup{m}, 0.25 \, \mu \textup{m}\right]$. \par %\footnote{{\bf For the given range of dust size distribution, the representative size $\sqrt{<a^2>}$ for the cross-section is 0.01 $\mu$m. We note that for the chemistry module, we use a size of 0.1 $\mu$m. Reduzing the monomer size by one order of magnitude would shift the dead-zone edge outwards by less then a factor of two, see section~\ref{sec:chemistry}.}} \citepads{1977ApJ...217..425M}.\par 
We verified that the particles couple to the gas by calculating the Stokes number. The Stokes number in the Epstein regime is 
\begin{equation}
\rm St= \frac{\rho_{dust} a}{\rho c_s} \Omega,
\end{equation}
with the orbital frequency $\Omega$, the sound speed $\rm c_s = \sqrt{\rm T k_b/(\rm \mu u)}$, the Boltzmann constant $\rm k_b$, the mean molecular weight of $\rm \mu = 2.353$, and the atomic mass unit u. The Epstein regime is valid if the particles are smaller than the mean free path of the gas molecule $a < \frac{9}{4} \lambda_{mfp}$. 
%For a density of $\rm \rho = 10^{-14} g/cm^3$ the mean free path of the gas molecule is $\lambda_{mfp}=\mu/(\sigma_{mol} \rho)=1.95 \cdot 10^{5} cm$ using a mean molecular weight of $\rm \mu=3.9 \times 10^{-24} g$ of a 5:1 $\rm H_2-He$ mixture and the cross section $\sigma_{mol}= 2 \times 10^{-15} cm^2$.
We estimated the Stokes number for our largest grain size $a_{max}=0.25\, \mu m$ by assuming a density of $\rm \rho = 10^{-14} g/cm^3$ and a temperature of 40K to $\rm St(0.25 \mu m) \propto 10^{-4}$. All the particles up to 0.25 $\mu m$ are very tightly coupled to the gas. For larger particles of about 100 $\rm \mu m$ we obtain a value of $\rm St(100 \mu m) \propto 0.04$. For this value we expect strong settling, radial migration, and particle concentration in particular in vortices \citep{cha10,meh12}. A local concentration paired with a higher emission efficiency of these particles in the (sub)mm range would lead to a higher brightness contrast
between the gap and its outer edges. 

\begin{table}[htbp]
\caption{Selected wavelengths for the simulated ALMA observations. \label{tbl:wavel}}
\begin{center}
\begin{tabular}{ccccc}
\hline\hline
ALMA band &  $\lambda$  &   $\nu$ &PWV & Octile\\
 & {[$ \rm \mu m$]} & {[$\rm GHz$]} & {[$\rm mm$]}   &   \\
\hline
9 & 441 & 679 & 0.472 $\pm$ 0.100 & 1st \\
7 & 871 & 344 & 0.658  $\pm$ 0.125 & 2nd \\
6 & 1303 & 230 & 1.262  $\pm$ 0.150 & 4th\\
\hline
\end{tabular}
\tablefoot{We consider the influence of thermal noise by precipitable water vapor (PWV) in the atmosphere using wavelength-dependent values listed in the table and the \textit{tsys-atm} option in the CASA simobserve task. Additionally, we include the influence of atmospheric phase noise into our calculations. 
We assume a phase screen blowing with a constant velocity above the ALMA array with the PWV values and deviations. In CASA this is numerically calculated with the \textit{sm.settrop} task in screen mode.}
\end{center}
\end{table}

\paragraph{The ALMA observations} were predicted on the basis of simulated thermal dust re-emission. ALMA operates in the wavelength range between $0.3$ and $9.3\, \rm mm$. We considered the complete 12m-array of 52 (sub)mm-antennas, which can be arranged in different configurations \citepads{2004AdSpR..34..555B}. We predicted ALMA observations with the CASA 4.2 simulator \citepads{2012ASPC..461..849P}. It calculates the visibilities in the u-v-plane as sampled by the array configuration, adds noise, and finally reconstructs the image with the CLEAN algorithm. The observing wavelengths listed in Table \ref{tbl:wavel} were selected to optimally suite the atmospheric windows in the (sub)mm range. The bandwidth for the (continuum) ALMA observations is $8\, \rm GHz$, the selected exposure time amounts to three hours. With the ALMA sensitivity calculator we estimated the sensitivity that can be achieved for selected wavelengths and spatial resolutions during this time. We considered the influence of thermal and phase noise as summarized in Table \ref{tbl:wavel}. The simulated observations were calculated for the reference position (max. height above horizon $\approx 45^\circ$) of the Butterfly Star (IRAS 04302+2247, $\alpha =$ 04h33m, $\delta = $ +22$^\circ$53$\arcmin$, J2000) in the Taurus-Auriga star-forming complex. We assumed the following distances of the object: $75\, \rm pc$ (e.g., V4046 Sgr \& TW Hya, \citealtads{2010ApJ...720.1684R} \& \citealtads{2006A&A...456..535S}), $100\, \rm pc$ (e.g., CQ Tau, \citealtads{2011A&A...529A.105G}), $120\, \rm pc$ (e.g., Oph IRS 48, \citealtads{van13} ),  and $140\, \rm pc$ (e.g., HD 169142, \citealtads{2012ApJ...752..143H}).
\subsection{Dead-zone edge versus a fully turbulent disk.} %FIX

In this section we compare the re-emission maps and simulated ALMA observations of our fiducial model \texttt{D2G\char`_e-2} including the dead-zone edge and model \texttt{D2G\char`_e-4}, which is fully turbulent. We focus on a single output after 410 inner orbits because a comparison with other snapshots taken during 100 additional inner orbits did not show any significant differences.\par

We explored which structures of the (dust) density distribution were conserved in the re-emission maps. Our goal is to identify the structures that are able to distingish between the two models. First we focus on ideal re-emission maps and then consider simulated observations with ALMA.\par

\begin{figure*}[t]
\psfig{figure=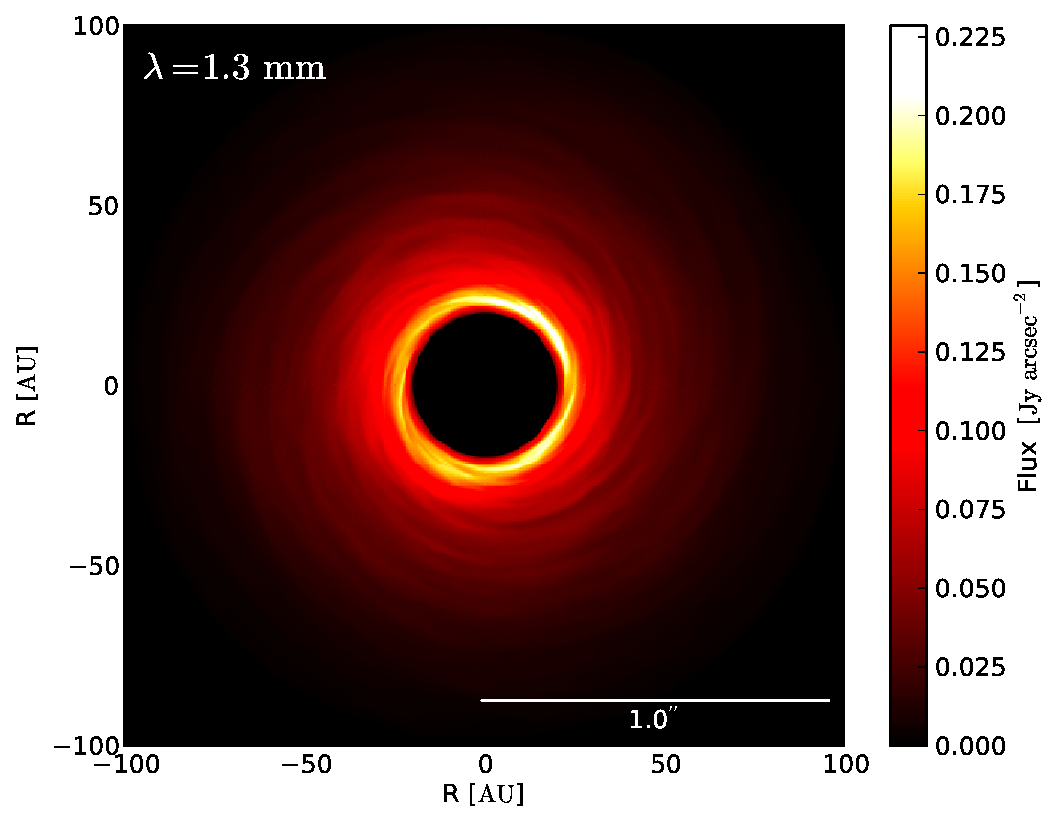,scale=0.5}
\psfig{figure=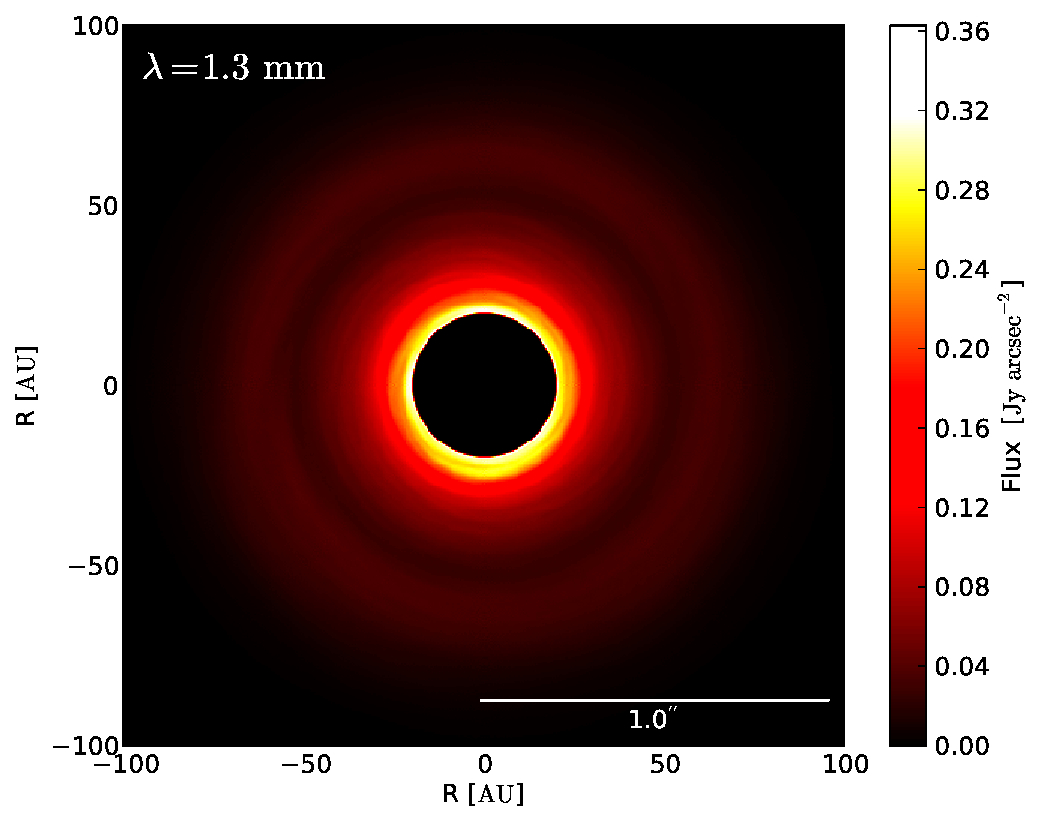,scale=0.5}\\
\caption{Selected re-emission maps for a wavelength of $1.3\, \rm mm$ and a distance of $100\, \rm pc$. The maps are calculated for model \texttt{D2G\char`_e-4} (left) and model \texttt{D2G\char`_e-2} (right). In these ideal re-emission maps, the turbulent structure in \texttt{D2G\char`_e-4} and the gap in \texttt{D2G\char`_e-2} are conserved.}
\label{fig:1.3mm-RM-fix}
\vspace{0.2cm}
\end{figure*}

For the model \texttt{D2G\char`_e-4} the turbulent structure of the density map is present in the re-emission map (see Fig. \ref{fig:1.3mm-RM-fix} left). 
It is very interesting that model \texttt{D2G\char`_e-2} shows a ring of reduced brightness at the exact position as the structure in the surface density map (see Fig. \ref{fig:1.3mm-RM-fix} right). This gap structure at the same time marks the edge of the dead zone. We note that the total re-emission fluxes are different between the models because of the different total dust mass as a consequence of the different dust--to--gas mass ratios. \par

\begin{figure}
\begin{minipage}{0.4\textwidth}
   \epsfig{figure=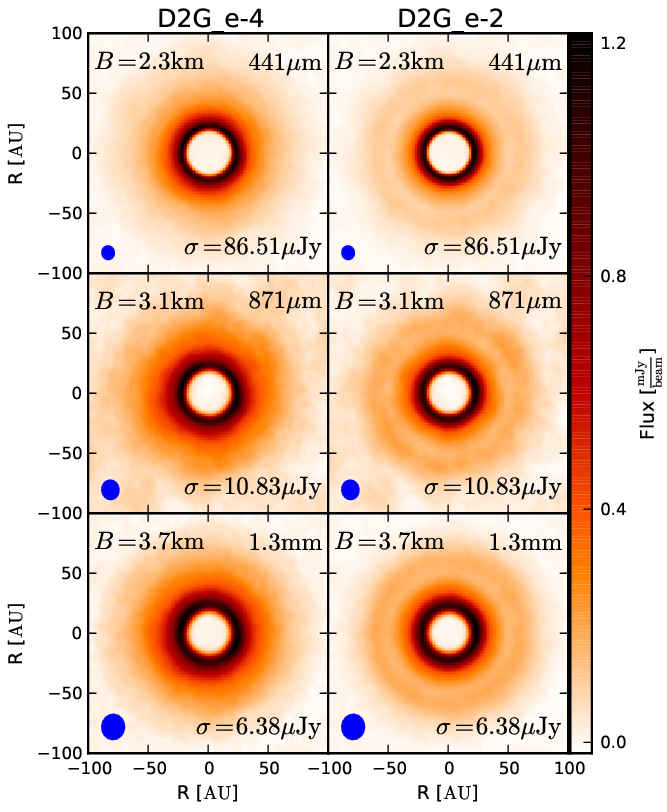,scale=0.8,clip=}
\end{minipage}
\caption{Selected simulated ALMA observations of the two disk models (left column: \texttt{D2G\char`_e-4}, right column: \texttt{D2G\char`_e-2}) at three different wavelengths (top: $441\, \rm \mu m$, center: $871\, \rm \mu m$, bottom: $1303\, \rm \mu m$) at a distance of $75\, \rm pc$. The longest baselines, synthesized beam sizes, and S/N are plotted into each panel. For comparison, the maps for model \texttt{D2G\char`_e-4} (left) are scaled with a factor 100.}
\label{fig:ALMA}
\end{figure}

\begin{figure}
\psfig{figure=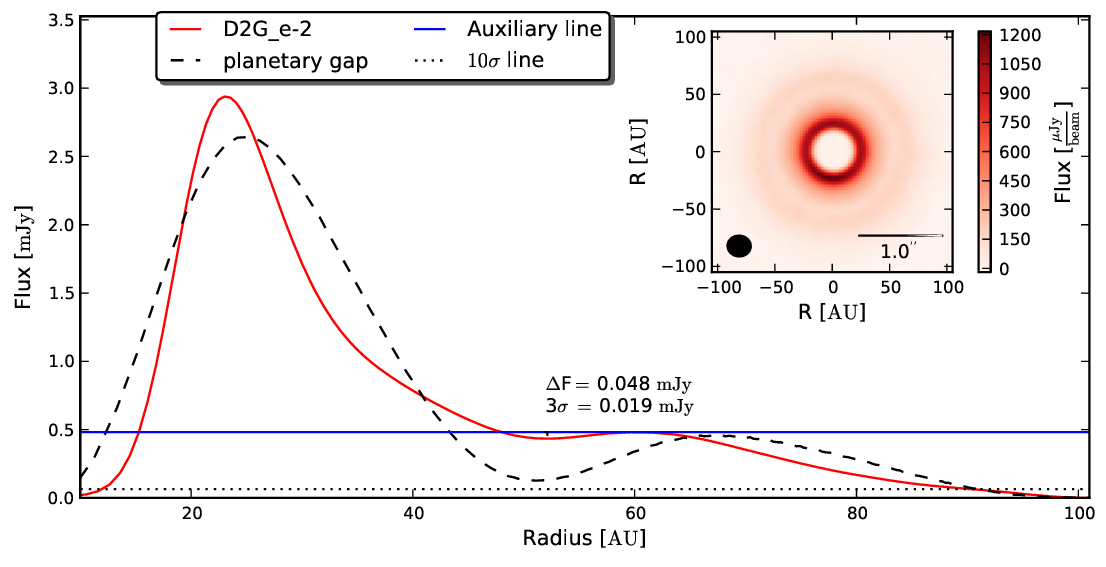,scale=0.45}
\caption{Radial brightness profile of a simulated ALMA map of model \texttt{D2G\char`_e-2} at a wavelength of $1.3\, \rm mm$ and a distance of $75\, \rm pc$. Compared to the local flux maximum at $60\, \rm AU$, the significance of the flux decrease at about $55\,\rm AU$ is  $7.7\sigma$. The dashed line shows the radial brightness profile of a disk that is perturbed by a giant planet with a planet-to-star mass ratio of $0.001$. The remaining disk parameters correspond to model \texttt{D2G\char`_e-2}. The $10\sigma$ level is indicated by the dotted line. The blue line is an auxiliary line that visualizes the maximum flux of the outer disk (red line) at any other location.}
\label{fig:radialBP}
\end{figure}

The next step was to probe whether it was feasible to trace the structures in these re-emission maps with ALMA. Fig. \ref{fig:ALMA} presents the simulated observations for the disk models \texttt{D2G\char`_e-2} and \texttt{D2G\char`_e-4} for an object at a distance of $75\, \rm pc$. Each plot includes the related synthesized beam size, the maximum baseline, and the signal-to-noise ratio $\sigma$. Furthermore, the figure illustrates the dependence of the observational results on the observing wavelength and atmospheric conditions. It clearly shows that it is possible to trace the gap located at the dead-zone edge in model \texttt{D2G\char`_e-2} with the distance- and wavelength-dependent significance listed in Table~\ref{tbl:gap_signi}. The value of the significance was derived by comparing the (maximum) signal at the outer edge of the gap and the signal at the gap center in the radial brightness profile (see Fig. \ref{fig:radialBP}).

\begin{table}[htbp]
\caption{Expected significance of gap detections in simulated ALMA observations of model \texttt{D2G\char`_e-2}. \label{tbl:gap_signi}}
\begin{center}
\begin{tabular}{l|llll}
\hline\hline
 $\lambda$ [$ \rm \mu m$] &  $75\, \rm pc$ &  $100\, \rm pc$ & $120\, \rm pc$ & $140\, \rm pc$\\

\hline
 441 	& $5.0\sigma$ & $3.9\sigma$  &	$2.6\sigma$ &	$2.2\sigma$\\	

 871 	& $11.7\sigma$ & $6.4\sigma$ &	$4.2\sigma$ &	$4.2\sigma$\\

 1303 	& $7.7\sigma$ & $4.8\sigma$ &	$3.3\sigma$ &	$2.4\sigma$\\
\end{tabular}
\end{center}
\end{table}
The $871\, \rm \mu m$ wavelength offers the potential of gap detections with the highest significance (see Table~\ref{tbl:gap_signi}). In particular, for objects at a distance of $140\, \rm pc$ a detection of a gap with $> 3\sigma$ is only possible at this wavelength. \par

\begin{table*}[t]
\caption{Expected significance of gap detections in simulated ALMA observations of the model \texttt{Net\char`_e2\char`_Pi} in $140\, \rm pc$ distance for different declinations $\delta$ of the object at a wavelength of $871\, \rm \mu m$. We select the configurations of the ALMA antennas with the optimum significance for each object position. $B_\text{max}$ indicates the longest baseline of the array configuration. \label{tbl:gap_signi_elo}}
\begin{center}
\begin{tabular}{l|llllll}
\hline\hline
 $\delta$ &  $23^\circ$ &  $13^\circ$ & $3^\circ$ & $-7^\circ$ & $-17^\circ$ & $-27^\circ$ \\

\hline

Significance 	& $5.9\sigma$ & $5.3\sigma$ &	$4.2\sigma$ &	$5.3\sigma$ & $4.5\sigma$& $4.1\sigma$\\

Beam size & $0.09^{^{\prime\prime}} \times 0.08^{\prime\prime}$&$0.09^{\prime\prime} \times 0.08^{\prime\prime}$&$0.09^{\prime\prime} \times 0.09^{\prime\prime}$&$0.10^{\prime\prime} \times 0.09^{\prime\prime}$&$0.09^{\prime\prime} \times 0.08^{\prime\prime}$&$0.11^{\prime\prime} \times 0.08^{\prime\prime}$\\

$B_\text{max}$ $[\rm km]$  & 2.7 & 2.7 & 2.7 & 3.1& 3.1 & 3.1\\

\end{tabular}
\end{center}
\end{table*}
Additionally, Table~\ref{tbl:gap_signi_elo} shows that the influence of the object position on the
significance of a gap detection is small as long as the object reaches a height of more than $40^\circ$ above the horizon.
For model \texttt{D2G\char`_e-4} the variations of the density are too low ($< 10\%$ at the midplane) and limited to a too small area ($<  1\times 1 \, \rm AU^2$) to be traced in the simulated ALMA observations. A characteristic ring structure such as the gap in the disk model \texttt{D2G\char`_e-2} is absent.\\
However, Fig. \ref{fig:radialBP} also shows that the considered high-angular submillimeter observations alone do not allow one to constrain the origin of the gaps. Giant planets potentially also open gaps in the disk density distribution \citepads{1979ApJ...233..857G,1984ApJ...285..818P}, which create a similar shape of the radial brightness profile, as an example case shows in Fig. \ref{fig:radialBP}. The disk with a planet is taken from \citetads{2013A&A...549A..97R} with a planet-to-star mass ratio of $0.001$. It also represents a simulated ALMA observation.\par
Furthermore, we produced scattered-light images of the two disk density models at a reference wavelength of $2.2\, \mu m$ (K band). They do not show any significant structure in the disk. It is similar to the corresponding scattered-light map of the initial disk setup. Therefore a simple 2D disk model is able to explain the scattered-light appearance of both models. The reason for this is the structure of the photosphere of the disk at the selected wavelength. The disk becomes optically thick for this wavelength already on a surface of the disk where the influence of turbulence and dead-zone edges is too low to be visible. A similar case has been investigated in detail for planet-induced gaps by \citet{rug14}.
\subsection{Consequences of turbulence for gas line observations}
\begin{figure*}
\begin{minipage}{0.4\textwidth}
\psfig{figure=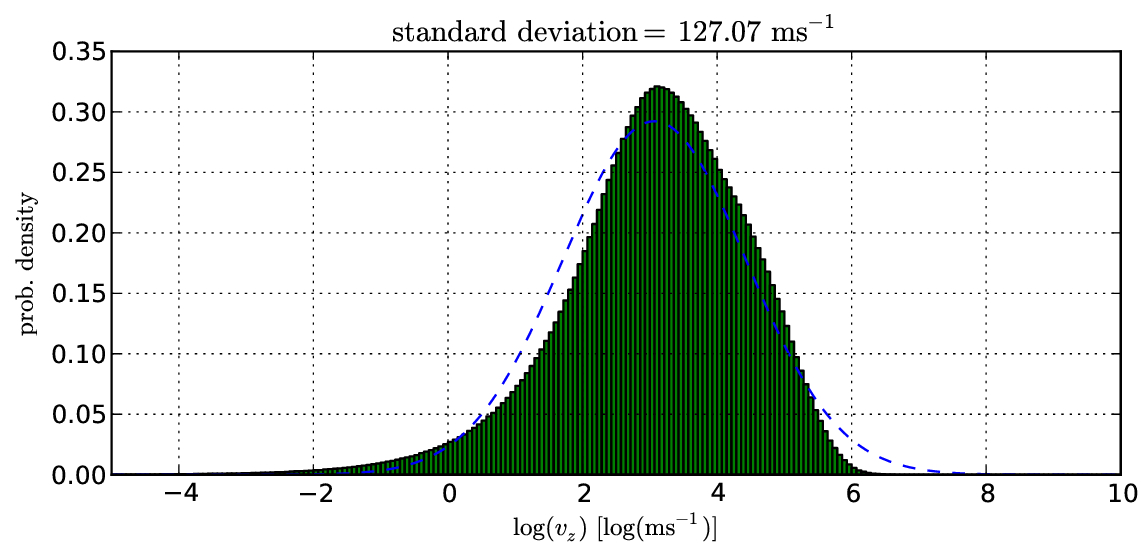,scale=0.55}
\psfig{figure=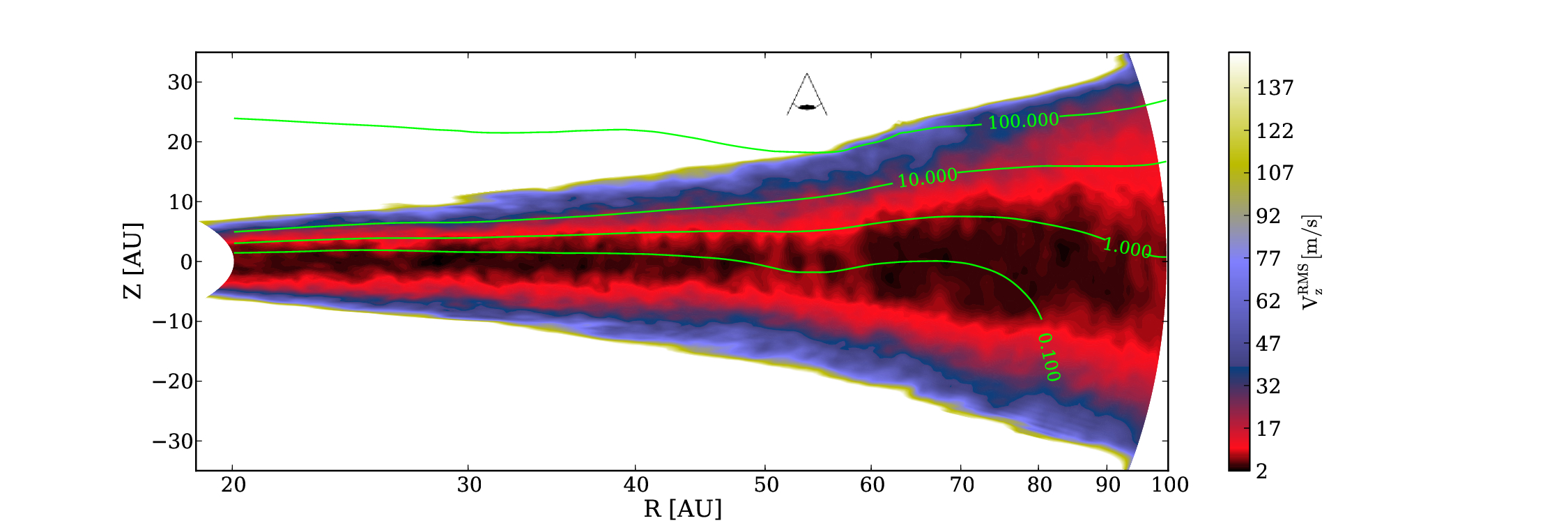,scale=0.45} 
\end{minipage}
\caption{Top: Histogram of the vertical velocity component in disk model \texttt{D2G\char`_e-4}. The mean value of the distribution is at zero and the FWHM is $\approx 127 \, \rm ms^{-1}$. The distribution is close to a log-normal distribution. Bottom: Contour plot of the vertical RMS velocity component in disk model \texttt{D2G\char`_e-2}. Dotted lines correspond to optical depth $\tau_z = 1$ lines along the vertical direction for an opacity of $\rm \kappa_{gas}=100,10,1,0.1, cm^2/g$. The line of sight of an observer is indicated by an eye sketch.}
\label{fig:vel_Z}
\end{figure*}

CO gas line observations of circumstellar disks frequently find broadened line profiles that cannot be explained by the thermal velocity of the molecules itself \citepads[e.g.,][]{1998A&A...339..467G,2007A&A...467..163P,2004ApJ...616L..11Q,gui12}.
Often, this broadening is considered to be the result of unresolved turbulence in the disk with a velocity component of the molecules toward or away from the observer. It is assumed that the absolute value of this velocity component is on the order of $100\, \rm ms^{-1}$ \citepads[e.g.,][]{2003A&A...399..773D,2007A&A...467..163P,2011ApJ...727...85H}. This value can be taken as the FWHM of the velocity distribution in the direction toward the observer. For the turbulent disk model \texttt{D2G\char`_e-4} with a face-on orientation we obtain $\Delta v_\text{z} \approx 127 \, \rm ms^{-1}$, see Fig.~\ref{fig:vel_Z}, top, using the full domain dataset. The velocity distribution is close to a log normal distribution, which is commonly expected for turbulence in this regime \citep{kri11,kri11b}. In the upper disk layers the velocity of the gas component is higher, which leads to a non-negligible influence of the tail end of the distribution \citep{sim11b}. The mean value of the Gaussian distribution confirms $0.0 \, \rm ms^{-1}$, as expected. In Fig.~\ref{fig:vel_Z}, bottom, we plot the turbulent component of the vertical velocity $v_z'$ for model \texttt{D2G\char`_e-2}, averaged over the azimuth, using a single time output after 410 inner orbits. The dotted lines represent the location of the optical depth unity $\tau_\textup{z} = \rm \int_{z}^{\infty} \kappa \cdot \rho dz=1$ along the vertical direction, with $\kappa=\tilde{\kappa} \cdot X$ and $\tilde{\kappa}$ being the line opacity of the molecule, while $X$ is the relative abundance of the molecule. The plot shows that line opacities higher than 10 cm$^2/g$ are more sensitive to the upper disk regions with strong turbulent velocity. For values below 0.1 cm$^2/g$ the optical depth would become optically thin. We note that to obtain an accurate estimate of a molecular line emission, the temperature, the dust properties and the ionization state has to be considered \citep{hen13,dut14}. We summarize that the turbulence in the two disk models is able to create line broadening. Especially the turbulent level of model \texttt{D2G\char`_e-4} fits currently observed magnitudes.
\section{Discussion}
In our models we considered a static resistivity profile. The change of surface density during the simulation will eventually drive a change of ionization of the gas. By including a dynamical resistivity, we expect an enhanced ionization in the gap, which could lead to an even stronger turbulent level in the gap. 
At the same time, a jump of surface density will decrease the ionization and consequently, the magnetic coupling. A similar study in unstratified simulations that includes the change of ionization in the bump and gap was described in \citet{joh11}. Here, they observed a strong pressure bump growing at the dead-zone outer edge. Future studies should include dynamical resistivity to verify the stability of the gap and bump structure.
\subsection{Convergence of the MRI}
In the appendix we have performed a resolution study of our models. We compare the results with simulations using half and a quarter of the resolution. The results show that the characteristic radial and vertical profiles of the turbulence converge well, especially for our fiducial model \texttt{D2G\char`_e-2}. We observe the emerging jump in surface density for resolutions down to 10 cells per H. The fully turbulent model \texttt{D2G\char`_e-4} shows a small increase of turbulence with lower resolution. A recent result that investigated the convergence of the MRI in local-box simulations reported by \citet{bod14} showed a non-convergence of the MRI in zero-net flux stratified disk simulations for resolutions of up to 200 cells per H. A similar result of non-convergence was found in unstratified simulations as well \citep{fro07}. Here, it was found that the convergence requires fixed diffusion terms \citep{fro07II}. The recent result for stratified simulations \citep{bod14} indicates that there might be similar effects. However, stratified simulations need to include the dynamo effect, which can generate a temporal small net flux that could sustain the turbulence \citep{dav10,gre10,flo12a}.  
Our global simulations are difficult to compare with high-resolution zero-net flux local-box simulations in the ideal MHD limit. First, we used a vertical net flux field in which convergence is found \citep{fro07,sim13}. Secondly, we used an explicit magnetic resistivity, which limits the growth of the MRI depending on the strength of the resistivity. We note that we might not be able to resolve very high magnetic Reynolds numbers, especially for values much higher than $\rm Re_m > 10^{4}$. The result of our resolution study of model \texttt{D2G\char`_e-4} could indicate that we still overestimate the level of the turbulence. \citet{flo12} showed a saturation of accretion stress for values above $\rm Re_m > 3000$ but this was tested for a zero-net magnetic field. There, a comparison with ideal MHD and non-ideal MHD simulations using the FARGO scheme showed an effective magnetic Reynolds number of around 5000 for a resolution of $\rm H/\delta x \sim 10$. Future simulations should investigate the convergence of the MRI for high magnetic Reynolds numbers in the low Prandtl number regime $\rm Pr=\nu/\eta$ using explicit diffusion terms.
%owever we don't expect a big change onto the large oder small scale structure. 
%\subsection{Non-linear Ohmic heating}
%In our simulations we assume linear Ohms law and the electric field on the local gas and its charged species is $\rm E=\frac{4 \pi \eta}{c}J$ with the current density $\rm J=\nabla \times B$. However, if the electric field is too strong the linear Ohms is not valid anymore. The critical electric field strength can be estimated with $\rm E \geq E_{crit}=\sqrt{\frac{6 m_e}{m_n}} \frac{k_B T}{{\it e} l_e}$ with the masses of electrons and neutral molecules $\rm m_e$ and $\rm m_n$, the Boltzmann constant $\rm k_B$, the elementary charge $e$ and the mean free path of electrons $l_e$ determined by electron-neutral collision \citep{lif81}. 
%An electric field larger than the critical value could enhance the magnetic diffusivity $\eta$ \citep{oku13b}. In our simulation we observe a electric field larger than the critical electric field, especially in the upper turbulent disk layers.
%\subsection{ambipolar and Hall diffusion}
\subsection{Ambipolar and Hall term}
In this work, we focused on the Ohmic diffusion term. The chosen disk model has a higher surface density than the minimum mass solar nebula (MMSN) model, therefore the Ohmic term is comparable with the ambipolar term at the midplane.
However, it is known that magnetic diffusion terms such as ambipolar and Hall diffusion can become the dominant diffusion terms especially in the upper layers of the disk \citep{war07,dzy13,tur14}. Recent simulations that included ambipolar diffusion \citep{bai13} and Hall diffusion \citep{les14,bai14} showed a strong laminar stress depending on the field geometry. A similar result, an increase of accretion stress when including Hall diffusion, was seen in multi-fluid global unstratified simulations by \citet{oke14}. In the outer regions, the MRI is expected to produce sufficient accretion stress depending on the net vertical magnetic field even with ambipolar diffusion \citep{sim13}. Recent non-ideal MHD local-box simulations in the outer disk regions by \citet{sim14} observed the generation of zonal flows in the regime dominated by ambipolar diffusion. %{\bf Considering ambipolar diffusion as dominant dissipation term, the detailed assumptions about the dust properties become  } {\bf We note that the detailed assumption about the dust properties of the dust size, the distribution and the porosity are less relevant, if we assume that ambipolar diffusion is the dominant disipation term.}
Future global stratified simulations should investigate the effect of other dissipation terms.

%\footnote{We note that the inner and outer regions,n close to the boundary (<5 AU), are influenced by the boundary condition and we do not include this region in our analysis.}
%We expect that by including all magnetic diffusion terms the dead-zone outer edge will be shifted in a region further out of the disk.
\section{Summary and conclusion}
We successfully combined non-ideal 3D global MHD simulations of the magneto--rotational instability (MRI) with 3D post-processing radiative transfer methods to predict synthetic ALMA maps.
Initial conditions of density and temperature were derived from best-fit models of a generic disk model, which was successfully applied for HH30, CB26, and the Butterfly star. The Ohmic resistivity was calculated consistently from a semi-analytical chemical module. Initial magnetic fields were vertical with a radial profile of 1/R and a field strength of 1 mGauss at 40 AU. We studied two models with different dust--to--gas mass ratios, $10^{-2}$ and $10^{-4}$. \\
%Additional simulations include particles with 10 different size bins (50 $\mu m$ up to 1 cm) to follow the particle motion at different levels of gas coupling.\\
\begin{itemize}
\item Both models developed a turbulent state driven by the MRI. The model with a dust--to--gas mass ratio of $10^{-2}$ included the outer dead-zone edge within the domain, having a laminar inner disk as well as a turbulent region at outer radii. The model with a reduced dust amount was fully turbulent with a time- and space-averaged accretion stress of around 0.01.  
\item Close to the dead-zone outer edge, we observed the formation of a large gap and bump structure in the surface density. This structure is similar to a zonal flow, axisymmetrically showing a ring of enhanced surface density. This gap and bump structure was formed due to the response of the MRI onto the change in the density and the Elsasser number. It was sustained during the simulation time of over 1000 inner orbits. The gap showed an increased MRI turbulence compared to the surroundings, with an Elsasser number of around 0.1. The dead-zone and bump structure are less turbulent, with Elsasser numbers below 0.01.  
\item The pressure bump at the inner edge of the surface density jump was strong enough to stop the radial drift of solid material. At this location the rotation velocity became slightly super-Keplerian and so prevented the radial migration of large particles.
\item Inside the surface density jump, we observed the formation of vortices by the Rossby wave instability with a lifetime of around 40 local orbits at a location of 60 AU. The vortices show a radial extent of two scale heights (~$\sim$~10~AU~) and an azimuthal extent of ten scale heights ($\sim$ 50 AU) with a strength of around -0.3 $[\Omega^{-1}]$.
%\item The simulations including different particle sizes reveal several ring structures as well as local concentrations of particles in both models. The ring structures with concentrated particles show radial separations length of the order of a pressure scale height.
\end{itemize}

The RT results and simulated ALMA observation showed the difficulty of distinguishing a turbulent from a nearly laminar disk. The turbulent structures in the gas and submicron dust are too small and are impossible to resolve. The same is true for specific time steps of the turbulent disk, which cannot be separated. In addition, there were no traces of turbulence or gaps in the scattered light maps of our models. We highlight that
\begin{itemize}
\item the gap and bump structure produced by the MRI at the dead zone outer edge can be traced by ALMA.
\item the fully turbulent disk model in face-on orientation is able to create line broadening of the observed size with a FWHM of 127 m$\rm s^{-1}$.
\end{itemize}
We note that a distinction between dead-zone edges and planetary gaps remains difficult. Further studies including a particle treatment are necessary to search for possible dust concentration and further asymmetries in disks all the more because our model predicts a concentration of larger particles close to the bump structure. In a follow-up work we will use the existing models and inject particles with different sizes during the simulation runtime by following the particle motion, we will study the radial drift and the concentration to verify with post-processing RT simulations whether it is feasible to trace the observed asymmetries in the disk.

\begin{acknowledgements}
We thank Sebastien Fromang for his great comments and contributions to this work.  
We also thank Andrea Mignone for his support regarding the PLUTO code and for providing the FARGO MHD method.
We thank Wladimir Lyra and Neal Turner for reading and giving comments on the manuscript.
Mario Flock is supported by the European Research Council under the European Union’s Seventh Framework Programme (FP7/2007-2013) / ERC Grant agreement nr. 258729. Parallel computations have been performed on the IRFU COAST cluster located at CEA IRFU.
J.P. Ruge acknowledges the financial support by the German Research Foundation (DFG WO 857/10-1). We thank F. Ober for helpful discussions regarding molecule line observations.
\end{acknowledgements}

\appendix
\section{Initial phase}
\begin{figure}
%\begin{minipage}{0.4\textwidth}
%\psfig{figure=PS/alpha-r.jpg,scale=0.40}
%\end{minipage}
%\hspace{-1cm}
\hspace{-1cm}
\begin{minipage}{0.4\textwidth}
\psfig{figure=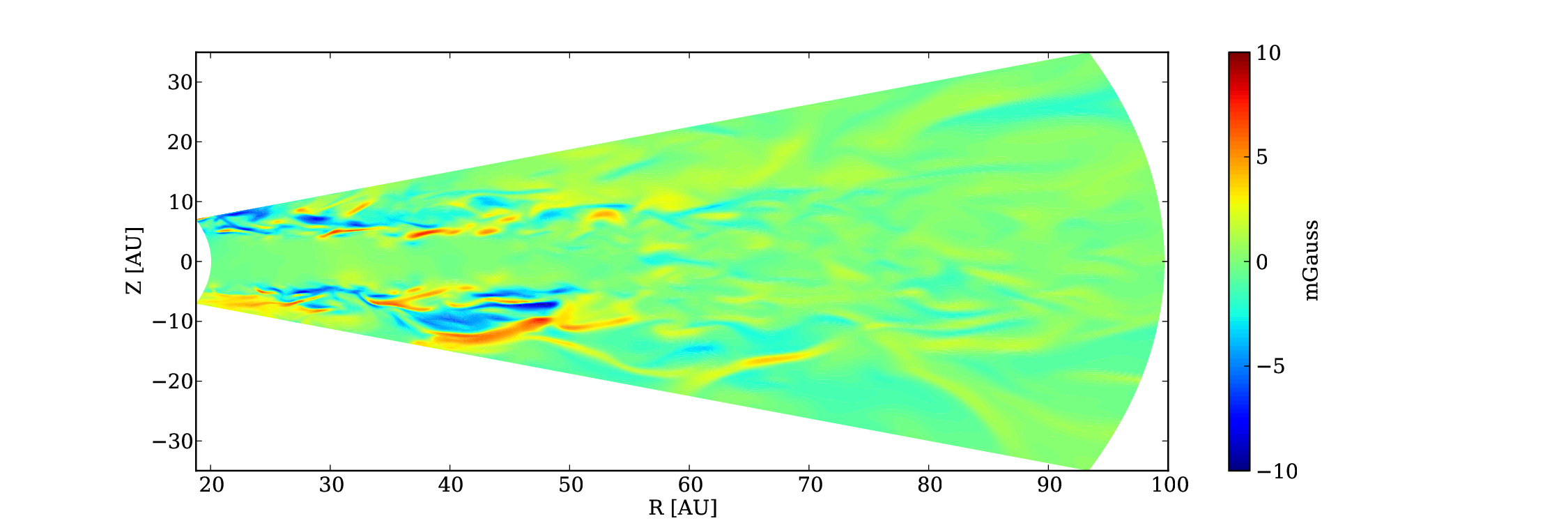,scale=0.30}
\psfig{figure=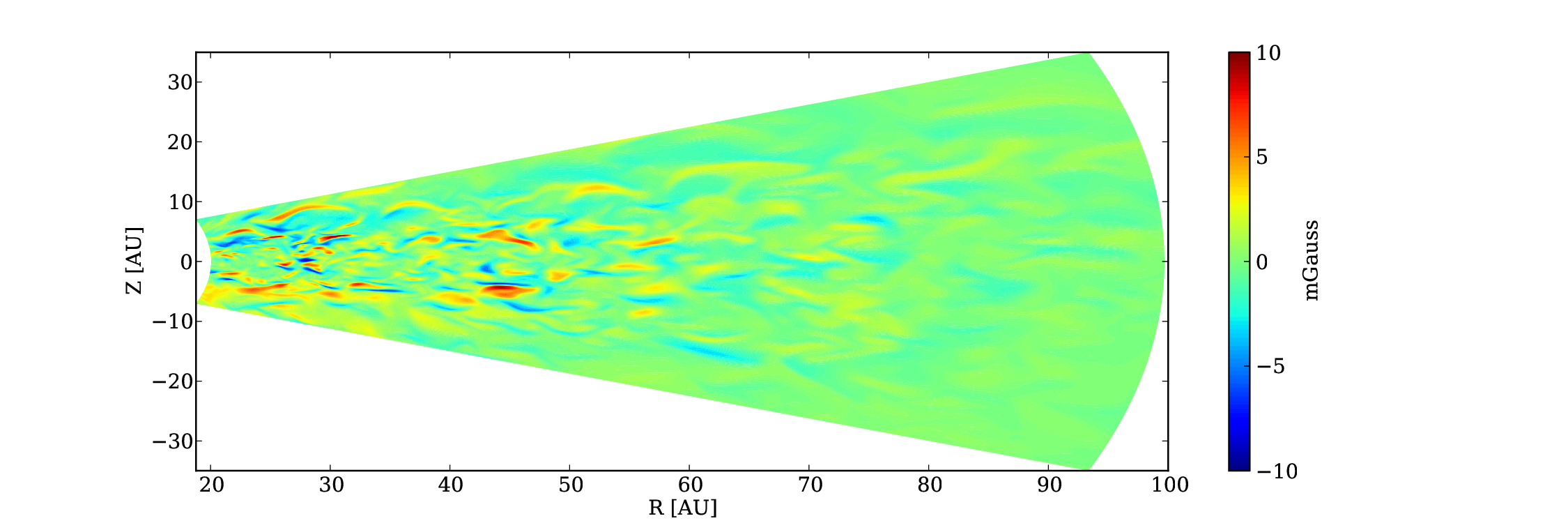,scale=0.30}
\end{minipage}
%\caption{Time evolution of the accretion stress $\alpha$ ({\it top left}) and time averaged radial $\alpha$ profile during the linear phase of MRI ({\it bottom left}). On the right, snapshots of the turbulent radial magnetic field in the $\rm R-Z$ plane after 100 inner orbits ($\sim$ 9000 years) for model \texttt{Net\char`_e2\char`_Pi} ({\it top right}), model \texttt{Net\char`_e4\char`_Pi}({\it middle right}) and \texttt{ZNet\char`_e4\char`_Pi} ({\it bottom right}).}
\caption{Snapshots of the turbulent radial magnetic field in the $\rm R-Z$ plane after 100 inner orbits for model \texttt{D2G\char`_e-2I} ({\it top}) and model \texttt{D2G\char`_e-4I} ({\it bottom}).}
\label{fig:apalpha}
\end{figure}

The model parameters are summarized in Table~\ref{tab:ap}. All runs show a strong linear MRI phase due to the net vertical field. 
The time--averaged $\alpha$ values during the first 300 inner orbits reach 0.038 for model \texttt{D2G\char`_e-2I} and 0.054 for model \texttt{D2G\char`_e-4I}, which will lead to a strong mass loss in both simulations.  
%In model \texttt{Net\char`_e2\char`_Pi}, the higher resistivity starts to damp the MRI and the $\alpha$ profile is reduced to the inner radius, ({\it black solid line}) whereas model \texttt{Net\char`_e4\char`_Pi} is fully turbulent. 
A snapshot of the turbulent radial magnetic field is shown in Fig. \ref{fig:apalpha} after 100 inner orbits for each model. 
The turbulent magnetic fields reach a strength of several mGauss. Model \texttt{D2G\char`_e-2I} shows a strongly damped inner midplane region,  Fig. \ref{fig:apalpha}, top. 
Between 50 and 60 AU a transition from this laminar dead-zone to a turbulent active zone is recognizable. We note that the midplane region is damped, while the upper layers are still active. 
%The models with an azimuthal extent of $180^\circ$ carry the keyword \texttt{Pi}. They are used for the initial runtime of 300 inner orbits to study the linear MRI phase and the saturation phase.
% Model \texttt{ZNet\char`_e4\char`_Pi} shows a reduced amplitude by roughly two orders of magnitude even having a reduced resistivity ({\it black dotted line}). \footnote{We note that we do not include the first 20 cells (3 AU) in our analysis. This region is influenced by the inner boundary condition (E.g. Fig. \ref{fig:alpha} ({\it dotted line}) the small increase of $\alpha$ in model \texttt{ZNet\char`_e4\char`_Pi}).}
%We summarize that a net magnetic flux is needed to trigger the MRI especially including magnetic dissipation. This result was already found in several local box simulations, recently again by \citet{les14,sim14,bai13}, and we can confirm this in global simulations. In the following chapters we focus on the net flux models, model \texttt{Net\char`_e4\char`_Pi} with a fully turbulent disk and model \texttt{Net\char`_e2\char`_Pi} having a transition of the dead-zone to the turbulent active zone. 
To continue the simulations, we reestablished the initial state of density and pressure and extended the azimuthal domain to full 2$\pi$.
We restarted the simulations from the initial conditions, but replaced the initial magnetic fields with the fields obtained after 200 inner orbits. This prevented the strong rise of turbulent activity while at the same time a turbulent steady state was reached quickly. The plasma beta $\rm \beta=2 P/B_z^2$ at 40 AU for the initial magnetic field was around 2200 (which corresponds to 1 mGauss), which led to a strong turbulent evolution. After 200 inner orbits, the large radial mass-loss also led to a reduction of the vertical magnetic flux. When we reset the density and pressure, the plasma beta for the vertical field at the midplane at 40 AU increased to around 44000.

\begin{table}
\caption{Model name, resolution, domain size, dust--to--gas mass ratio, and averaged accretion stress.\label{tab:ap}}
\begin{tabular}{lllll}
Model& $\rm N_r x N_\theta x N_\phi$ & $\rm \Delta r : \Delta \theta : \Delta \phi$ & D2G & $<\alpha>$\\  \hline
\hline
\texttt{D2G\char`_e-4I}  & $256$x$128$x$256$ & 20-100 : 0.72 : $\pi$ & $10^{-4}$ & 0.054\\ 
\texttt{D2G\char`_e-2I}  & $256$x$128$x$256$ & 20-100 : 0.72 : $\pi$ & $10^{-2}$ & 0.038 \\ 
\end{tabular}
\end{table}

\section{Resolution study}
\begin{figure}
%\begin{minipage}{0.4\textwidth}
%\psfig{figure=PS/alpha-r.jpg,scale=0.40}
%\end{minipage}
%\hspace{-1cm}
\begin{minipage}{0.4\textwidth}
  \epsfig{figure=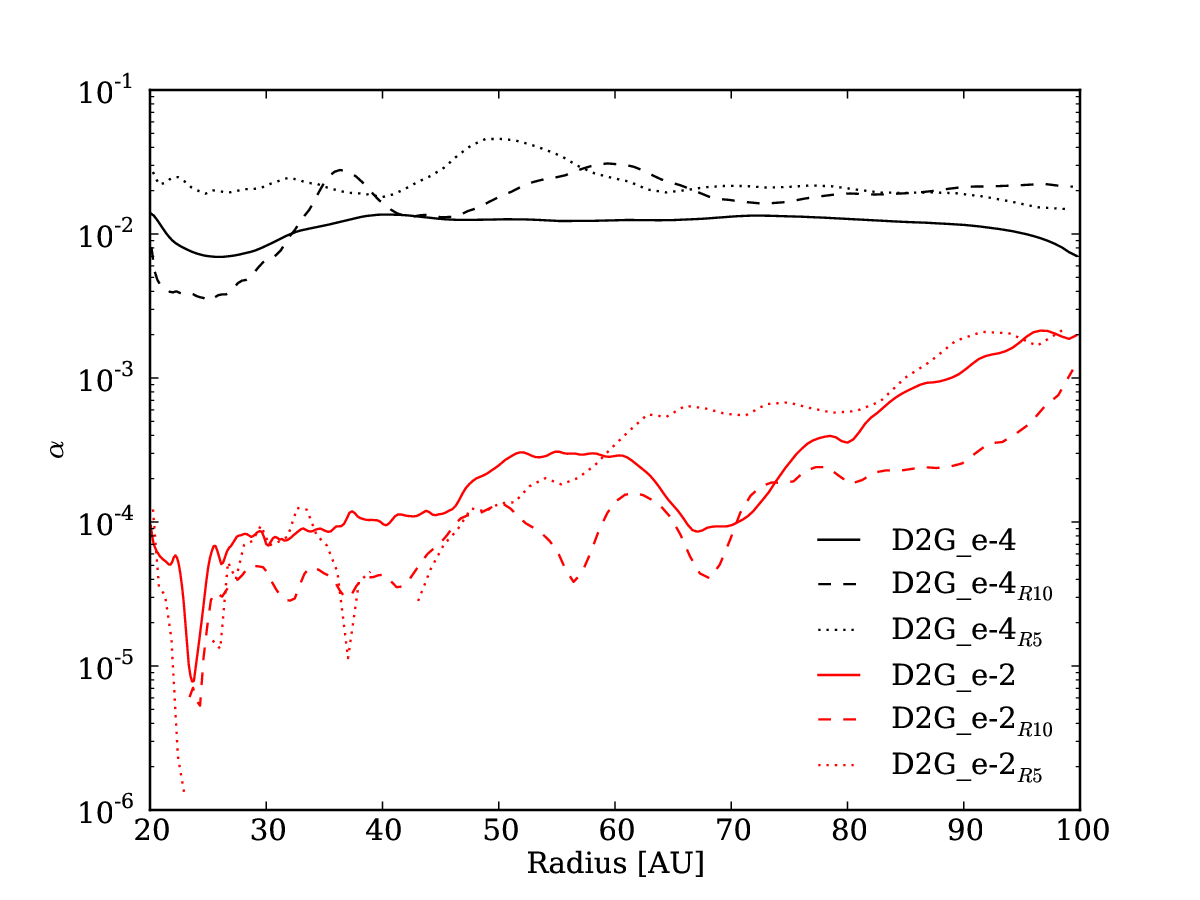,scale=0.45}
  \epsfig{figure=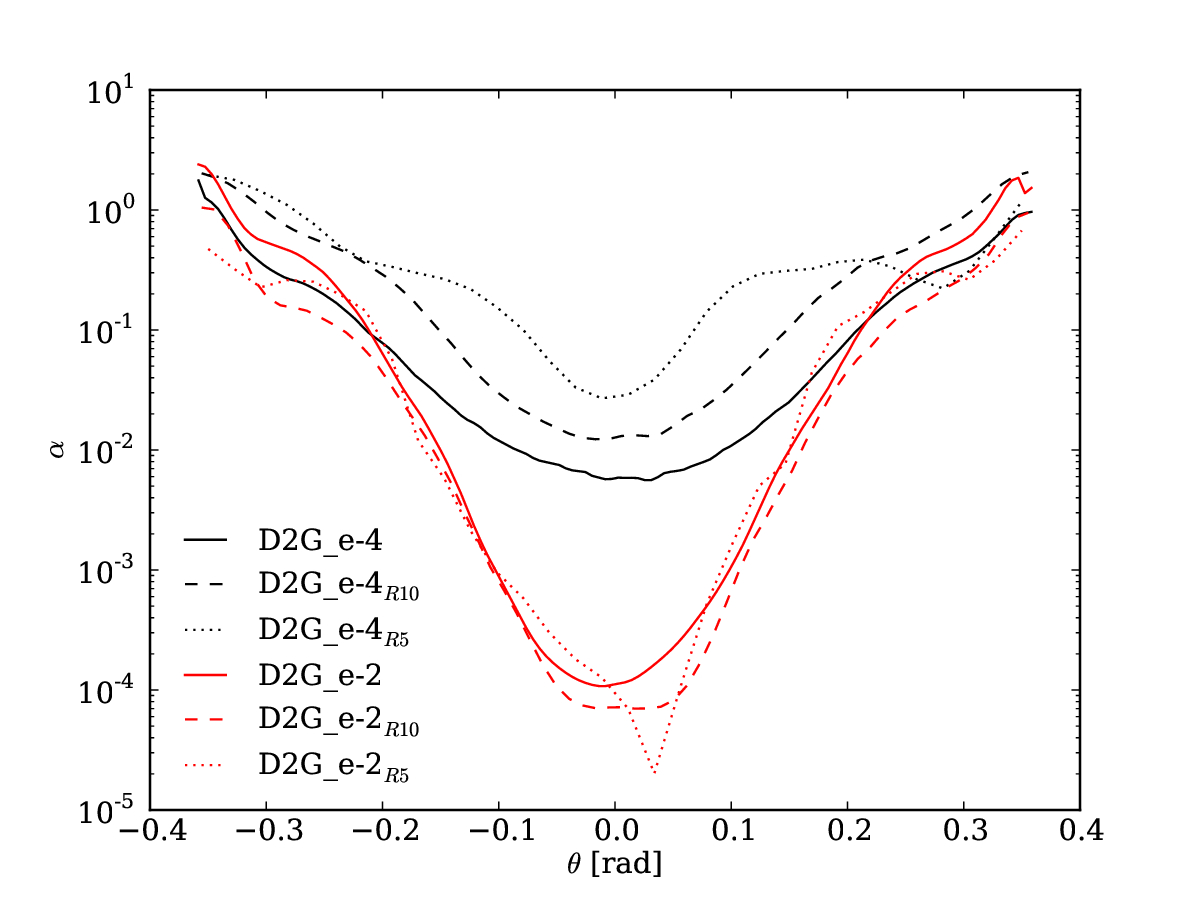,scale=0.45}
\end{minipage}
\caption{Radial (top) and vertical (bottom) profile of the accretion stress $\alpha$ for three different resolutions and two different dust--to--gas mass ratios averaged over 200 inner orbits (300 to 500 inner orbits). Here, the cylindrical height Z is given by $\rm Z=R_{mid} \sin{\theta}$ with $\rm R_{mid}=60AU$.}
\label{fig:res_comp}
\end{figure}

In the following section, we perform a resolution study to determine the robustness of our results. 
We included four additional simulations, two for each dust--to--gas mass ratio, with half and a quarter of the resolution.
%The results for the accretion stress are summarized in Table.~\ref{tab:apb}.
In Figure~\ref{fig:res_comp} we compare the radial and vertical profiles of accretion stress for all different models, see Table~\ref{tab:mod_res}. To calculate the results, we used the same time average, between 300 and 500 inner orbits. We used exactly the same restart technique for all models. The vertical profiles were taken from the middle of the domain. 
Our fiducial disk model with a dust--to--gas mass ratio of $10^{-2}$ shows a very good convergence even with only five grid cells per scale height. This result was surprising because we expected to resolve the fastest growing mode of the MRI with eight cells per H or more \citep{flo10}. This could indicate that the FARGO MHD method is able to resolve the fastest growing modes with slightly fewer cells. The models with a reduced dust amount show a small decrease in the accretion stress with higher resolution. We explain this by the fact that only the largest and strongest modes are resolved in the low-resolution runs. In general, small-scale turbulence can have a dampening effect on the larger scales. 
We observe the emerging jump in surface density for resolutions down to 10 cells per H. The lowest resolution has around five cells per H, and this model is unable to develope the gap and bump structure. In the following, we check for the low-resolution model \texttt{D2G\char`_e-2$_{R10}$} how the gap and jump structure appears when observed with ALMA.

\begin{table}
\caption{Top: Model name, resolution, and grid cells per H \label{tab:mod_res}}
\begin{tabular}{lll}
Model name & $\rm N_r x N_\theta x N_\phi$ & $\sim$ H/$\Delta$x \\  \hline
\hline
\texttt{D2G\char`_e-4}  & $256$x$128$x$512$ & 20 \\ 
\texttt{D2G\char`_e-2}  & $256$x$128$x$512$ & 20 \\ 
\hline
\texttt{D2G\char`_e-4$_{R10}$}  & $128$x$64$x$256$ & 10 \\ 
\texttt{D2G\char`_e-2$_{R10}$}  & $128$x$64$x$256$ & 10 \\ 
\hline
\texttt{D2G\char`_e-4$_{R5}$}  & $64$x$32$x$128$ & 5 \\ 
\texttt{D2G\char`_e-2$_{R5}$}  & $64$x$32$x$128$ & 5 \\ 
\end{tabular}
\label{tab:apb}
\end{table}

% \footnotemark
%\footnote{{\bf The cylindrical height is given by $\rm z=R_{mid} \sin{\theta}$ with $\rm R_{mid}=60AU$}}
%
\begin{figure}
\begin{minipage}{0.4\textwidth}
  \epsfig{figure=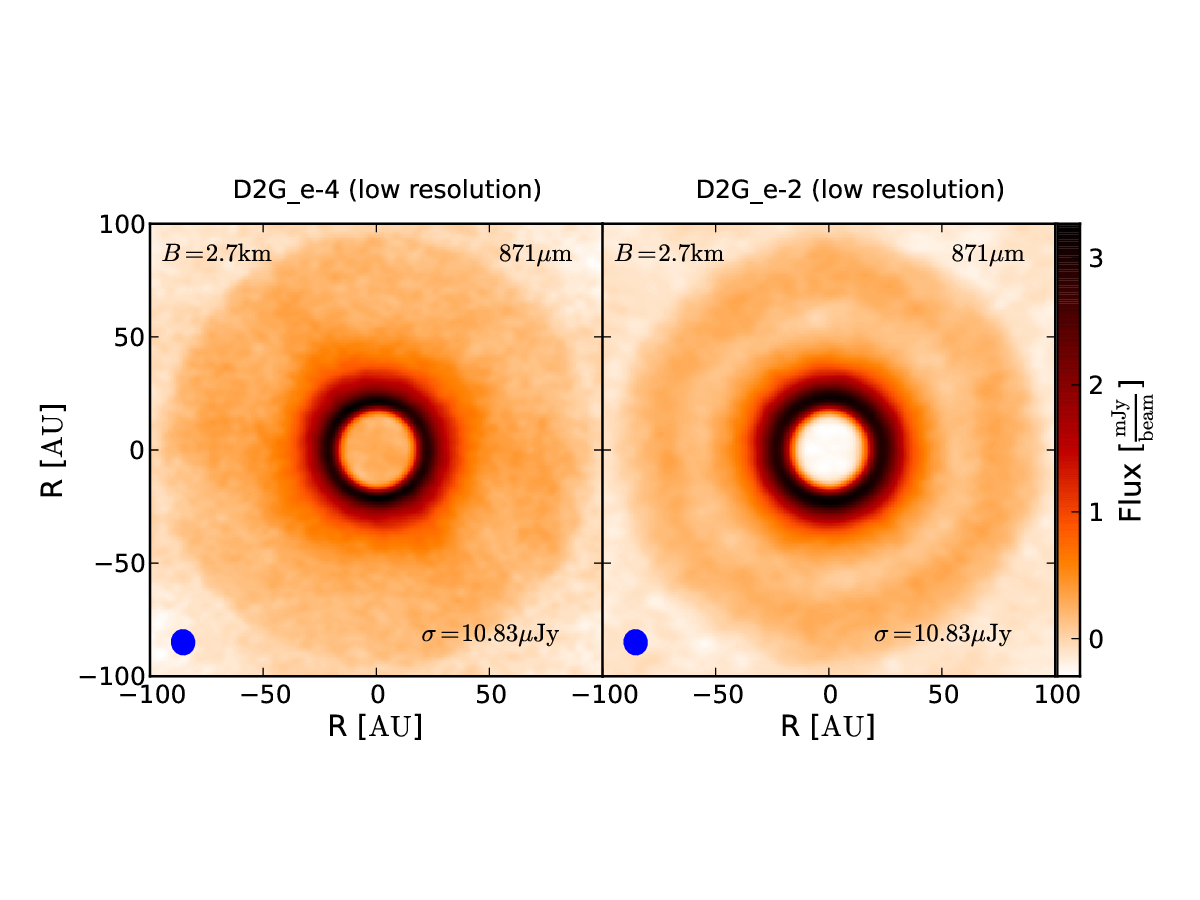,scale=0.45}
  \epsfig{figure=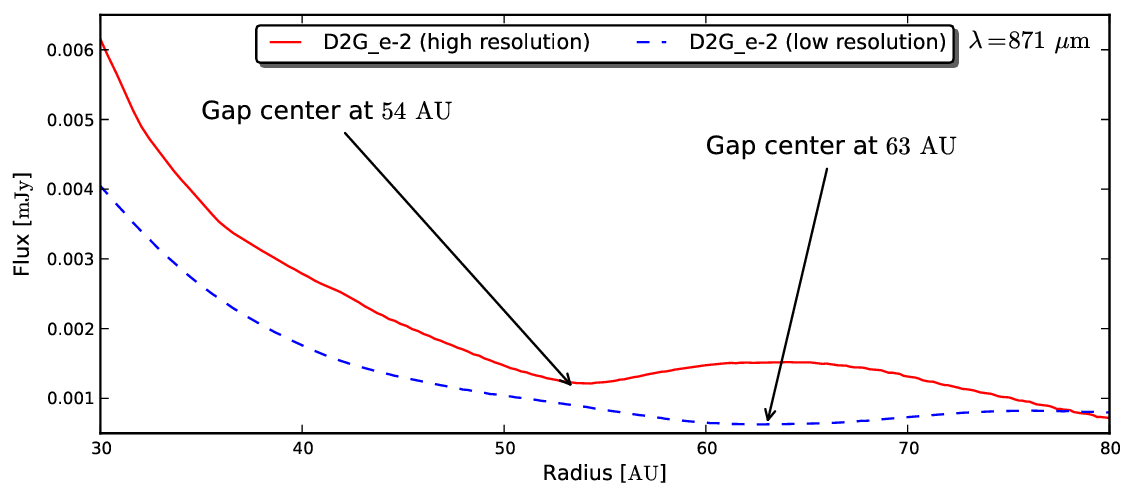,scale=0.45}
\end{minipage}
\caption{Top: Selected simulated ALMA observations of the two disk models (left column: \texttt{D2G\char`_e-4$_{R10}$}, right column: \texttt{D2G\char`_e-2$_{R10}$} at a wavelength of $871\, \rm \mu m$ and a distance of $75\, \rm pc$. 
For comparison, the maps for model \texttt{D2G\char`_e-4$_{R10}$} (left) are again scaled with a factor 100. Bottom: Radial brightness profile of a simulated ALMA map of the two models.}
\label{fig:apb}
\end{figure}

%\subsection{Radiative transfer results of model }
The radiative transfer results for the models of reduced resolution achieve the same precision in calculating of the temperature
distribution of the disk and the resulting re-emission maps as the high-resolution maps (see chapter~\ref{sec:rt}). For the disk model \texttt{D2G\char`_e-4}
it is not possible to distinguish the radiative transfer results
for the different resolutions of the MHD simulation. We present this result
through a simulated ALMA observation shown in Fig. \ref{fig:apb} (top left),
which is the counter piece to the ALMA simulation of the high-resolution
case shown in Fig. \ref{fig:ALMA} (mid-left). The re-emission maps for the
disk simulation \texttt{D2G\char`_e-2$_{R10}$} again show a ring of reduced brightness
at the exact position of this structure in the surface density map.
In the radial brightness profile as well as in the surface density map, the location of the gap is about $10\, \rm AU$ more distant from the disk center than for the high-resolution map (see Fig. \ref{fig:apb}, bottom).
This gap is detectable with ALMA; for example, we present a simulated ALMA
observation at a wavelength of $871\, \rm \mu m$ for a disk at $75\, \rm pc$ distance in Fig. \ref{fig:apb}, top.
Finally, ALMA is able to measure the position and width of the gap on the order of the beam size.

\bibliographystyle{aa}
\bibliography{flock}
\end{document}